\RequirePackage[2020-02-02]{latexrelease}
\documentclass[twocolumn]{aastex63}
\turnoffeditone

\usepackage{amsmath}	% to enable cases environment

\newcommand{\eps}{{\rm erg\,s^{-1}}}
\newcommand{\epcs}{{\rm erg\,cm^{-2}\,s^{-1}}}
\newcommand{\epc}{{\rm erg\,cm^{-2}}}
\newcommand{\Xb}{\ensuremath{\overline{X}}}
\newcommand{\xbar}{\ensuremath{\overline{X}}}
\newcommand{\mns}{M_{\rm NS}}
\newcommand{\rns}{R_{\rm NS}}
\newcommand{\qnuc}{Q_{\rm nuc}}
\newcommand{\yign}{y_{\rm ign}}
\newcommand{\zcno}{Z_{\rm CNO}}
\newcommand{\xte}{{\it RXTE}}

\newcommand{\igr}{{\it INTEGRAL}}

\newcommand{\opz}{1.259}

\submitjournal{ApJS}

\shorttitle{Burst model-observation comparisons}
\shortauthors{Galloway et al.}

\begin{document}

\title{Robust inference of neutron-star parameters from thermonuclear burst 
observations}

\newcommand*{\MSU}{Department of Physics and Astronomy, Michigan State University, East Lansing, MI 48824, USA}
\newcommand*{\JINA}{Joint Institute for Nuclear Astrophysics-Center for the Evolution of the Elements, Michigan State University, East Lansing, MI 48824, USA}

\correspondingauthor{Duncan K. Galloway}
\email{duncan.galloway@monash.edu}

\author[0000-0002-6558-5121]{Duncan K. Galloway}

\affil{School of Physics \& Astronomy,
Monash University, Clayton, VIC 3800, Australia}

\affil{OzGRav-Monash, School of Physics \& Astronomy, 
Monash University, Clayton, VIC 3800, Australia}

\affil{Institute for Globally Distributed Open Research and Education (IGDORE)}

\author[0000-0003-4023-4488]{Zac Johnston}
\affil{\MSU}
\affil{\JINA}

\author[0000-0003-3441-8299]{Adelle Goodwin}
\affil{International Centre for Radio Astronomy -- Curtin University, GPO Box U1987, Perth, WA 6845, Australia}

\author[0000-0002-2332-8178]{Chong-Chong He}
\affil{Department of Astronomy, University of Maryland, College Park, MD 20742-2421, USA}

\begin{abstract}

Thermonuclear (type-I) bursts arise from unstable ignition of accumulated fuel on the surface of neutron stars in low-mass X-ray binaries. Measurements of burst properties in principle enable observers to infer the properties of the host neutron star and mass donors, but a number of confounding astrophysical effects contribute to systematic uncertainties. Here we describe some commonly-used approaches for determining system parameters, including composition of the burst fuel, and introduce a new suite of software tools, {\sc concord}, intended to fully account for astrophysical uncertainties. Comparison of observed burst properties with the predictions of numerical models is a complementary method of constraining host properties, and the tools presented here are intended to make comprehensive model-observation comparisons straightforward. When combined with the extensive samples of burst observations accumulated by X-ray observatories, these software tools will provide a step-change in the amount of information that can be inferred about typical burst sources.
\end{abstract}


\section{Introduction} \label{sec:intro}

Thermonuclear (type-I) bursts are observed from neutron stars accreting from low-mass ($\lesssim 1\ M_\odot$) binary companions \cite[e.g.][]{gal21a}. Many thousands of events have been detected from over a hundred known sources \cite[e.g.][]{minbar}, and these events have emerged as an important way to determine the properties of the host neutron-star (NS) systems. The presence of bursts clearly indicates a NS host, and the measured burst properties can help constrain the source distance \cite[]{kuul03a}, 
accreted fuel composition \cite[e.g.][]{gal03d}, and even the NS mass and radius \cite[e.g.][]{ozel16b}. Comparisons with numerical models may also provide constraints on the key nuclear reactions which shape the burst lightcurves \cite[e.g.][]{meisel18a}, with a high degree of complementarity to nuclear experiments.

However, much attention over recent years has focussed on the systematic issues which may affect inferred quantities of the NS hosts \cite[e.g.][]{slb10,kajava14}. These systematic errors arise from astrophysical effects such as non-Planckian burst spectra, and anisotropic emission \edit1{arising from the disk geometry}.
\edit1{Additionally there is evidence that energetic bursts may temporarily disrupt the accretion flow \cite[e.g.][]{zand11a,worpel15,deg18a}, which further complicates interpretation of the observations.
Fully accounting for such effects} is challenging, because we generally lack independent measurements of the interesting system parameters. 

One approach is to use 
thermonuclear burst simulations, which are available over a range of complexity \cite[e.g.][]{gal21a}. In some cases, simple one-dimensional/one-zone codes which use an analytic approach to calculate the ignition conditions (based on the accretion rate and fuel composition) may provide reasonable fidelity \cite[e.g.][]{gal06c,goodwin19c}.
To fully simulate the bursts, time-dependent codes are used, and multiple (vertical) zones are typically required to fully resolve the thermal and compositional profile \cite[e.g. {\sc kepler};][]{woos04}.
The principal input parameters include the accretion rate (per unit area), $\dot{m}$; the composition of the accreted fuel, usually quantified as $(X_0, Z_{\rm CNO})$ where $X_0$ is the mass fraction of hydrogen, and $Z_{\rm CNO}$ the mass fraction of CNO nuclei, which drive the hot-CNO cycle burning between bursts; and (typically) a parameter describing the degree of heating from below the model domain, usually labeled as ``base flux'' $Q_b$.
To date comparisons of burst observations and models have not, we argue, been fully realised, due to both a dearth of suitable high-quality data, as well as suitable software tools for the comparison.

Here we explore the ways that host system  parameters affect the observed properties of thermonuclear bursts, and \edit1{provide a set of numerical routines called {\sc concord} to correct for them}.
In \S\ref{sec:data} and sub-sections, we describe the typical characteristics of burst observations along with those properties that can be 
inferred from simple calculations. % or by comparison with numerical models.
In \S\ref{subsec:code} we describe the basic operating principles of the {\sc concord} code.
In \S\ref{subsec:anisotropy} we describe the important issue of anisotropy of the X-ray emission  from bursting NSs, and its treatment in the code.
We  describe how the distance and accreted composition may be inferred in \S\ref{subsec:distance} and \S\ref{subsec:energetics}, respectively.
In \S\ref{subsec:simobs} we describe the set of model simulations used to verify the calculations in the preceding section.
In \S\ref{sec:analysis} we \edit1{give examples of} a number of methods to infer properties of the bursting source, based on the extent of available observational data, and assess their accuracy.
In \S\ref{subsec:validation} we apply these methods to a range of simulated data to assess what degree of systematic errors these measurements might be subject to.
In \S\ref{subsec:fuelcomp}--\S\ref{subsec:zerobursts} 
we demonstrate the utility of the code via applications to various different cases, depending upon the availability of observations.
In \S\ref{sec:disc} we summarize our results and discuss the implications.

\section{Deriving burst properties with concord}
\label{sec:data}

\label{subsec:data}

The fundamental observable 
is the time-history of the burst flux $F_b$, the lightcurve $(t_i,F_{i})$. Here the $F_{i}$ may be in instrumental units of count~s$^{-1}$ (or perhaps count~s$^{-1}$~cm$^{-2}$), or may be estimates of the bolometric flux between times $t_i$ and $t_i+\Delta t_i$, derived from spectral model fits of the spectrum 
over that interval.
The net (excluding the persistent emission) burst spectra generally can be well fit with a blackbody model, and it is generally assumed that such models can provide a reasonable estimate of the bolometric flux 

Provided the $F_i$ are in units of flux, or the flux can be estimated from the intensity,  we may measure the peak flux, $F_{\rm pk}$, and integrate to estimate the burst fluence, $E_b$. 
Where the bursts are very short, or observed with low signal-to-noise, a spectrum extracted over an interval covering the entire event may substitute for a fluence measurement.

Given one or more bursts, we can derive constraints on the recurrence time $\Delta t$. A pair of bursts separated by an interval with uninterrupted coverage by X-ray instruments offer an unambiguous measurement of the recurrence time, but such measurements are rare due to the typically low (a few \%) average duty cycle for most instruments \cite[e.g.][]{minbar}.

\edit1{In the next sections we describe how these observables may be used with the {\sc concord} code to constrain burst properties.}

\begin{deluxetable*}{lccl}
\tablecaption{Key burst parameters and associated {\tt concord} functions or methods\label{tab:keyparams}}
\tablehead{
\colhead{Parameter} & \colhead{Symbol} & \colhead{Units} & \colhead{Associated function} }
\startdata
Burst flux\tablenotemark{a} & $F_b$ & $\epcs$ & \\
Burst peak flux\tablenotemark{a} & $F_{\rm pk}$ & $\epcs$ & \\
Burst fluence\tablenotemark{a} & $E_b$ & $\eps$ & \\
Persistent flux\tablenotemark{a} & $F_p$ & $\epcs$ & {\tt fper} \\
Bolometric correction & $c_{\rm bol}$ & \nodata & \\
Burst recurrence time\tablenotemark{a} & $\Delta t$ & hr & \\
Average burst rate & $\overline{R}$ & hr$^{-1}$ & {\tt tdel\_dist}\\
Burst/persistent emission anisotropy & $\xi_{b,p}$ & \nodata & {\tt diskmodel.anisotropy} \\
Ratio of accretion to thermonuclear energy\tablenotemark{a} & $\alpha$ & \nodata & {\tt alpha} \\
Burst/persistent luminosity & $L_{b,p}$ & $\eps$ & {\tt luminosity} \\
Eddington luminosity & $L_{\rm Edd,\infty}$ & $\eps$ & {\tt L\_Edd} \\
System inclination & $i$ & deg &\\
System distance & $d$ & kpc & {\tt dist} \\
Neutron star mass & $M_{\rm NS}$ & $M_\odot$ & {\tt calc\_mr} \\
Neutron star radius & $R_{\rm NS}$ & km & {\tt calc\_mr} \\
Surface gravity & $g$ & $10^{14}\ {\rm g\,cm^{-2}}$ & {\tt g}\\
Surface redshift & $1+z$ & \nodata & {\tt redshift} \\
Accretion rate per unit area & $\dot{m}$ & g~cm$^{-2}$ & {\tt mdot} \\
Accreted hydrogen mass fraction & $X_0$ & \nodata & {\tt X\_0}\\
Mean hydrogen mass fraction at ignition & $\Xb$ & \nodata & {\tt hfrac} \\
Metallicity & $Z_{\rm CNO}$ & \nodata & \\
Base flux & $Q_b$ & MeV/nucleon & \\
Burst energy generation & $Q_{\rm nuc}$ & MeV/nucleon & {\tt qnuc} \\
Accretion energy generation & $Q_{\rm grav}$ & MeV/nucleon & \\
Burst ignition column & $y$ & g~cm$^{-2}$ & {\tt yign} \\
\enddata
\tablenotetext{a}{These quantities we consider the fundamental observables for quantifying burst properties, although they may also be derived from parameters used for input into models via the listed function/method (e.g. $F_p$ from $\dot{m}$ via equation \ref{eq:fper} and the {\tt fper} method)}
\end{deluxetable*}

\subsection{Code architecture}
\label{subsec:code}

The code is provided as a set of Python modules available on GitHub\footnote{\url{https://github.com/outs1der/concord}; \cite{concord_v1}}. There are three principal components; the functions in {\tt utils.py}, the anisotropy treatment in {\tt diskmodel.py} and the observed and model burst classes in {\tt burstclass.py}.

The functions in {\tt utils.py} provide the basic functionality 
for constraining burst properties (Table \ref{tab:keyparams}) and propagating uncertainties. 
We adopt a Monte-Carlo (MC) approach via the {\sc astropy} {\tt Distribution} package \cite[]{astropy13,astropy18} to accommodate a wide range of probability distribution functions (PDFs) for the measured quantities or system parameters. Input values (including measured quantities) can be provided in four different ways: scalar, value with symmetric error, value with asymmetric error, or an arbitrary distribution array. In the first three cases, quantities with error estimates will be converted to a {\tt Distribution} object of user-defined size \edit1{via the {\tt value\_to\_dist} function}, and with a symmetric or asymmetric normal distribution. 
\edit1{In the case of asymmetric errors, the distribution is constructed assuming the first quantity corresponds to the maximum likelihood for the parameter distribution, following \cite{barlow03}. If the provided statistics are instead cumulative (i.e. 50/16/84th percentile, for $1\sigma$) the user can invoke {\tt value\_to\_dist} with the {\tt statistics="cumulative"} flag to provide the correct shape.}
The fourth case allows an arbitrary array to be passed to the routines, whether it be a set of alternative possible values or a synthetic distribution derived from some other calculation.
These {\tt Distribution} objects can then be used in calculations as for scalars, hence providing uncertainty propagation at the cost of additional computation. 
Input values can be provided with units, or if units are absent, will be assumed to have the standard units for MINBAR quantities \cite[]{minbar}; for flux, $10^{-9}\ \epcs$, burst fluence $10^{-6}\ \epc$, recurrence time hr, and so on.

A key objective of the code is to include of the effects of anisotropic emission, which is thought to affect both the burst and persistent flux (e.g. \citealt{he16}; see also section \ref{subsec:anisotropy}).
All functions can operate assuming isotropic distributions for the burst or persistent emission ({\tt isotropic=True}), but by default will include the possible effects of anisotropic emission via the {\tt diskmodel.py} routine. This routine incorporates the modelling of \cite{he16}, via ASCII tables provided with the code.
The user can also directly run the simulation code \textsc{DiskAnisotropy}\footnote{\url{https://github.com/chongchonghe/DiskAnisotropy}} to generate tables for other disk models beyond 
those included. The following parameters can be specified: disk outer radius, disk inner radius (optionally allowing a gap between NS surface and disk), disk inner height, disk inclination, and gravitational radius of the NS for the incorporation of GR effects. The columns of the output tables are: the line-of-sight inclination angle, 
the (inverse) direct and reflected anisotropy factors $\xi_d^{-1}$ and $\xi_r^{-1}$ (with $\xi_b^{-1}=\xi_d^{-1}+\xi_r^{-1}$), and the persistent anisotropy factor $\xi_p^{-1}$.
The user can provide the inclination value (or a distribution), or a uniform distribution can be generated automatically between user-defined ranges (defaulting to 0--$75^\circ$, appropriate for non-dipping sources).

The outputs from each function depend on the type of input; for scalar inputs without uncertainties the isotropic calculation will return a single value. More typically, where uncertainties are included, or a range of inclinations is provided, the functions will return a tuple giving the central (\edit1{50th percentile}) value and the upper and lower 68\% uncertainties.
With the {\tt fulldist=True} option the functions will return a dictionary including distributions for the results, as well as the intermediate values, providing a complete picture of the calculation for plotting or further calculations.
The size of the generated distributions is inherited from any input distributions provided, or can be set explicitly with the {\tt nsamp} parameter.

Where observed or model burst lightcurves are available, the classes in {\tt burstclass.py} provide a way to represent those observations and perform various standard analyses, including observation-model comparisons. The {\tt ObservedBurst} class can be instantiated from an ASCII file giving the burst flux as a function of time, or in a number of other ways. The reference bursts provided by \cite[]{gal17a} can be read in directly, provided the data is available %\todo{somewhere....}
locally.
An example {\tt KeplerBurst} model burst class is provided, which offers a number of ways of reading in {\sc kepler} burst runs, and which can be adapted to outputs of different codes.

\subsection{Emission anisotropy}
\label{subsec:anisotropy}

One of the most significant astrophysical uncertainties that has largely been neglected to date is the anisotropy of the burst (and persistent) emission.
The persistent emission is thought to arise largely in a boundary layer where the accretion disk flow meets the NS \cite[e.g.][]{done07}. In contrast, the burst emission is thought to arise more or less uniformly over the NS surface, once the accreted fuel has spread to cover it, and ignited. In both cases the influence of the surrounding accretion disk and binary companion produces a range of intensities depending upon the system inclination to the observer's line of sight. 

\edit1{Correcting for anisotropy is further complicated by the observation of two main spectral states for low-mass X-ray binaries (LMXBs),  characteristically  ``hard'' or ``soft'', which are thought to indicate two quite different geometries of the accretion flow. The hard (or ``island''; cf. with \citealt{hvdk89}) state is associated with low accretion rates, and is inferred to result from a thin disk possibly truncated above the NS surface, with accretion occurring mainly via a spherically-symmetric, optically thin flow. The soft (or ``banana'') state is thought to indicate accretion primarily through a boundary layer where the disk meets the NS surface, and also where most of the accretion luminosity is released. }

There have been several attempts to model the effect on the burst and persistent emission \cite[e.g.][]{ls85,fuji88}, the most recent in the context of high ($>1$) reflection fractions inferred from the emission from \edit1{NSs with puffed up accretion disks} \cite[]{he16}.
This effect can be significant, up to a factor of two for the burst emission, and even higher for persistent flux.

In the {\sc concord} code we adopt the treatment of \cite{he16}, which offers a number of different models based on the modelling performed by those authors, via the {\tt diskmodel} sub-module.
The burst (persistent) anisotropy factor $\xi_b$ ($\xi_p$) introduced 
is defined in the same sense as \cite{fuji88}, such that the total burst (persistent) luminosity is
\begin{equation}
L_{b,p}=4\pi d^2\xi_{b,p} F_{b,p}.
\end{equation}
The range of values of the anisotropy factors can thus be understood that if $\xi<1$ the flux is enhanced (i.e. preferentially beamed) toward our line of sight (so that $L_b$ would be {\it overestimated} were 
the anisotropy
not included in the calculation), while if $\xi>1$ the flux is suppressed.
Since the distribution of the burst flux over the NS is likely different from that of the persistent emission, we define a different anisotropy factor $\xi_p$ relating the observed persistent flux $F_p$ to the total luminosity. 
The {\tt anisotropy} function calculates both anisotropy parameters for a given inclination $i$ (or distribution thereof); this function is integrated into many of the other functions in the repository, such that the user can simply specify the desired inclination range to take these factors into account.

\begin{figure}[ht!]
\epsscale{1.2} % added for preprint version
\plotone{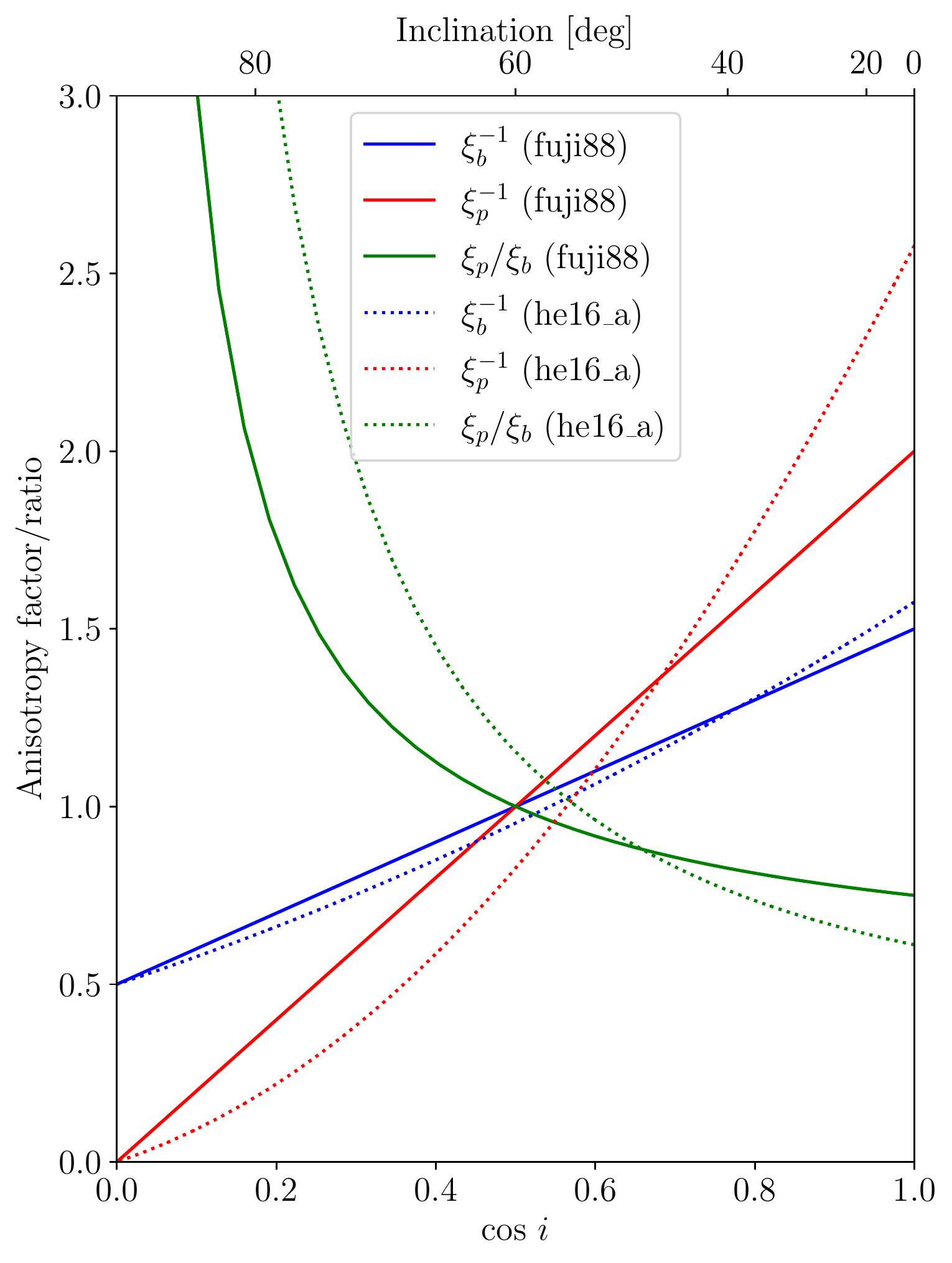}
\caption{Anisotropy factors $\xi_b$, $\xi_p$ affecting the burst and persistent emission (respectively), predicted by \cite{fuji88} and \cite{he16} (labeled as ``fuji88'' and ``he16\_a'', respectively) and adopted by the {\sc concord} routines.
The two treatements generally differ in their predictions  by less than 10\% over most of the range of inclination values, but a larger discrepancy is found for the ratio $\xi_p/\xi_b$, which enters into the calculation of burst energetics via the $\alpha$-parameter (equation \ref{eq:alphath}).
\label{fig:anisotropy}}
\end{figure}

The modelling of \cite{he16} predicts $\xi_b$, $\xi_p$ as a function of system inclination for a given disk geometry (Fig. \ref{fig:anisotropy}). For comparison, we also plot the predictions of \cite{fuji88}, which are qualitatively similar. At low inclinations (corresponding to a system observed ``face-on''; $\cos i \rightarrow 1$) it is predicted that both the burst and persistent emission are preferentially beamed towards the observer, by a factor of up to 2.5 (2.0 for \citealt{fuji88}) for the persistent emission, or 1.5 for the burst.
At high inclinations (observing ``edge-on'', $\cos i \rightarrow 0$) the emission will become increasingly suppressed by the accretion disk, dropping to half the isotropic value for the burst emission, and vanishing entirely for the persistent.
The system inclination for LMXBs is notoriously difficult to measure. For non-dipping sources, the likely range is up to $72^\circ$, while dipping sources are likely 
$i\gtrsim75^\circ$ 
(\citealt{parmar86}; see also \citealt{gal16a}).
For an isotropically-distributed population of non-dipping sources, the ratio $\xi_b/\xi_p$ (which enters into the calculation for the $\alpha$-value; see \S\ref{subsec:energetics}) is roughly uniformly-distributed between 0.53--1.64, rising a little towards the high end. The median value is 1.13, while the 95\% confidence interval is 0.73--1.49.

These model predictions must be viewed with caution, as they cannot be independently verified.
\edit1{Given the expected variation in the disk geometries in LMXBs in response to accretion rate, ideally the user should choose a disk geometry appropriate for the observed spectral state of the target source at the time of the burst observation. The default adopted by {\sc concord}, model (a) of \cite{he16}, is a compromise between the two characteristic disk geometries, as it is flat but extends to the NS surface. For this reason, it may not be strictly correct for either state, but offers reasonable consistency with the earlier work of \cite{fuji88}.}
New anisotropy models may be added to the {\tt diskmodel} module of {\sc concord}, and selected with the {\tt model} parameter.

\subsection{Burster distances}
\label{subsec:distance}

A significant fraction of observed bursts exhibit characteristic variations in their spectral hardness (or the blackbody temperature and normalisation) around their maximum that indicate the presence of photospheric radius-expansion (PRE; \citealt{lew84}). 
Even if such variations are not present, 
the peak flux $F_{\rm pk}$ provides constraints on the source distance, since the maximum burst luminosity is limited to (roughly) the Eddington luminosity \cite[e.g.][]{lew93}:
\begin{eqnarray}
  L_{\rm Edd,\infty} & = & \frac{8\pi G m_{\rm p} M_{\rm NS} c
  [1+(\alpha_{\rm T}T_{\rm e})^{0.86}]} {\sigma_{\rm T}(1+X)[1+z(R)]} % \xi_b
       \nonumber \\
  & = & 2.80\times10^{38} \left(\frac{M_{\rm NS}}{1.4M_\odot}\right)
 \frac{1+(\alpha_{\rm T}T_{\rm e})^{0.86}}{(1+X)}
\nonumber \\ & & \times\  % line break
    \left[\frac{1+z(R)}{\opz}\right]^{-1}\ % \xi_b\
              \eps .
  \label{ledd}
\end{eqnarray}
Here
$T_{\rm e}$ is the
temperature of the \edit1{scattering electrons}, $\alpha_{\rm T}$ parametrizes the temperature dependence of the electron scattering
opacity 
\cite[$=1/4.5\times10^8$~K$^{-1}$;][]{pac83a,pac83b},
$m_{\rm p}$ the proton mass, $\sigma_{\rm T}$ the Thomson 
cross-section, and $X$ the
mass fraction of hydrogen in the atmosphere.
The final factor in
square brackets represents the gravitational redshift 
at the photosphere $1+z(R)=(1-2GM_{\rm NS}/R
c^2)^{-1/2}=1.259$ for $M_{\rm NS}=1.4\ M_\odot$ and $R=R_{\rm NS}=11.2$~km.
The effective redshift measured at the peak of a PRE burst may be lower than the value at the NS surface, while the photosphere is expanded during the radius expansion episode (i.e. $R\ge R_{\rm NS}$).

By equating the maximum flux of these events with the Eddington luminosity, the distance to the source can be estimated as
\begin{eqnarray}
 d & = & \left(\frac{L_{\rm Edd,\infty}}{4\pi \xi_b F_{\rm pk, RE}}\right)^{1/2} \nonumber \\
   & = & 8.83
	\left( \frac{\xi_b F_{\rm pk, RE}}{3\times10^{-8}\ \epcs} \right)^{-1/2}
       	\left(\frac{M_{\rm NS}}{1.4M_\odot}\right)^{1/2}
\nonumber \\ & & \times\  % line break
	\left[\frac{1+z(R)}{\opz}\right]^{-1/2}
	(1+X)^{-1/2}\ {\rm kpc}.
 \label{disteq}
\end{eqnarray}
Here the $\xi_b$ factor accounts for the possible anisotropy of the burst emission
(see \S\ref{subsec:anisotropy}).
We note that this  expression omits the \edit1{electron} temperature dependence of equation \ref{ledd}; one could in principle include this \edit1{term} in the \edit1{distance} calculation, 
\edit1{or more precise corrections to the opacity \cite[e.g.][]{suleimanov11a,poutanen17}; but such corrections are not usually warranted given the precision achievable for typical distance measurements.}

The source distance can be estimated from the peak flux of a PRE burst based on equation \ref{disteq}, using the {\tt dist} method 
in {\sc concord}.
If the theoretical Eddington luminosity 
is adopted ({\tt empirical=False}),
values for $M_{\rm NS}$, $T_e$, $X$ and $1+z$ are assumed; these values, or distributions thereof, can also be provided by the user. No intrinsic uncertainty is otherwise assigned to the theoretical value,
but the observer must take into account the uncertainties in the various parameters that go into this calculation. 

As is obvious from equation \ref{disteq}, several factors will contribute to statistical (and possibly systematic) uncertainties in distance estimates derived in this manner. First, the NS mass $M_{\rm NS}$ is typically unknown for burst sources, although has been constrained in some cases based on the evolution of the X-ray spectrum in the burst tail (e.g. \citealt{ozel16a}, although see also \citealt{poutanen14}).
Second, the radius $R$ at which the redshift should be calculated may not be clear from the observations. In principle one could take the inferred blackbody normalisation at the time of maximum flux, but for most PRE bursts the maximum flux is achieved close to the ``touchdown'' point, where the photosphere has returned (more or less) to the NS surface \cite[e.g.][]{gal06a}.
Third, the 
precise value for $X$, the hydrogen fraction at the height in the atmosphere where the Eddington luminosity is reached; this quantity may be effectively zero for most bursts \cite[e.g.][]{gal06a,bult19b}, but some uncertainty remains.
Fourth, even if the inclination $i$ of the system is known, converting to the anisotropy factor likely requires modelling of the effect of the disk on the radiation field arising from the burst \cite[e.g.][]{he16}.

These issues may be avoided by adopting instead the empirical Eddington luminosity measured by \edit1{\citealt{kuul03a} (setting {\tt empirical=True} for the {\tt dist} method)}, for a sample of bursters with independently-measured distances from their host globular clusters. The mean value of $(3.79 \pm 0.15)\times 10^{38}\ \eps$ is somewhat difficult to reconcile with the theoretical expectation in equation \ref{ledd}, even for the most extreme values of the relevant quantities which affect it.
If we adopt the empirical Eddington luminosity of \cite{kuul03a}, the only remaining factor is the burst emission anisotropy $\xi_b$. 
Since the empirical value effectively averages over the anisotropy of each source in the sample (weighted implicitly by the number of bursts from each) it may be argued that this effect should not be included. However the user can make their own decision and/or explore the consequences with or without the additional anisotropy correction.

For either choice of Eddington limit,
the approach adopted in {\sc concord} to take the emission anisotropy into account is to generate a distribution of $\xi_b$ values, based on an isotropic distribution of inclinations within user-defined limits. The presence (absence) of X-ray dips occurring at the binary orbital period will constrain the inclination to greater (less) than $75^\circ$ 
\cite[e.g.][]{parmar86}. However, more stringent constraints may be available based on modelling of optical orbital variations (for example).

\subsection{Burst energetics and fuel composition}
\label{subsec:energetics}

\label{subsec:energy}

The fluence (and the shape of the burst profile) is determined by the amount of accumulated fuel and its composition. For accretion of mixed H/He with hydrogen mass fraction $X_0$, the composition at ignition may be modified substantially by $\beta$-limited CNO burning, ``catalysed'' by CNO nuclei, with mass fraction $Z_{\rm CNO}$. In extreme cases the recurrence time is sufficiently long that the accreted hydrogen at the base of the fuel layer is exhausted, and ignition of intense, short He-rich bursts occur.

The burst ignition column can be estimated from the fluence $E_b$ as
\begin{eqnarray}
y & = & \frac{L_b (1+z)}{4\pi R_{\rm NS}^2 Q_{\rm nuc}}\ 
                                                            \nonumber \\
  & = & 1.91\times10^8 \left(\frac{\xi_b E_{\rm b}}{10^{-6}\ \epc}\right)
                      \left(\frac{d}{10\ {\rm kpc}}\right)^2\ 
\nonumber \\ & & \times\  % line break
             \left(\frac{Q_{\rm nuc}}{5.22\ {\rm MeV/nucleon}}\right)^{-1}
                      \left(\frac{1+z}{\opz}\right)
\nonumber \\ & & \times\  % line break
                      \left(\frac{R_{\rm NS}}{11.2\ {\rm km}}\right)^{-2}
                      {\rm g\,cm^{-2}},
\label{eq:column}
\end{eqnarray}
where $\qnuc$ is the energy generation rate for the burst.
This quantity can be calculated for a given burst with measured fluence $E_b$ and known distance $d$ and $\qnuc$, using the {\tt yign} function.

For a fuel layer consisting of mixed H/He, $\qnuc$ depends on the mean hydrogen fraction at ignition, \Xb. 
In the absence of other information, $\qnuc$ (and hence $y$) has in the past been estimated by adopting a value for 
$\Xb$,
and using the relation 
$Q_{\rm nuc}=1.6 + 4\Xb$~MeV/nucleon 
\cite[e.g.][and references therein]{gal03d}. This expression includes $\approx35$\% losses
attributed to neutrino emission \cite[]{fuji87}. However, the 35\% value applies only to $\beta$-decays in the much more extensive {\it rp}-process burning chain.
Recent studies with the 1-D numerical code {\sc kepler} suggest instead that 
\begin{eqnarray}
Q_{\rm nuc} &= & 1.31+6.95\Xb-1.92\Xb^2 \nonumber \\
& \approx & 1.35 + 6.05 \Xb
\label{eq:qnuc}
\end{eqnarray}
\cite[]{goodwin19a}.
For fuel with solar composition (i.e. $X=0.7$), $Q_{\rm nuc}=5.23$~MeV/nucleon.
The average H-fraction at ignition $\Xb$ is a lower limit for the H-fraction in the accreted fuel, $X_0$, as for most sources the H-fraction will be reduced by steady burning prior to burst ignition.
The $\qnuc$ values can be calculated from $\Xb$ either with the old relation  or either of the two more recent approximations, using the {\tt qnuc} function.

Provided the burst recurrence time is known, $\qnuc$ can also be estimated via the $\alpha$-parameter, the (observed) ratio of the burst to persistent flux.
From data taken prior to the burst ignition we can measure the persistent flux $F_{\rm per}$;
this quantity is expected to be related to the \edit1{mass accretion rate per unit area} $\dot{m}$, as follows: % gal03d, equation 2
\begin{equation}
F_{\rm per,\infty} = \frac{L_{\rm per}}{4\pi d^2} = 
\frac{R_{\rm NS}^2 \dot{m} Q_{\rm grav}}{d^2 (1+z)\xi_p c_{\rm bol}},
\label{eq:fper}
\end{equation}
where $Q_{\rm grav} = c^2z/(1+z) \approx GM_{\rm NS}/R_{\rm NS}$ is the gravitational energy release per gram. 
The bolometric correction $c_{\rm bol}$ accounts for the experimental limitation that the persistent flux can only be measured over a limited instrumental passband. The bolometric correction is the ratio of the estimated bolometric flux to the band-limited value.
The {\tt fper} function implements equation \ref{eq:fper} for the case of model bursts simulated at specific values of $\dot{m}$, given the input distance $d$ and any anisotropy constraint.

The persistent spectra of burst sources are  more complex than the bursts themselves, and typically exhibit thermal as well as non-thermal components. % [{\bf ref}].
These parameters may be combined to estimate the $\alpha$-value, the ratio of accretion to thermonuclear burning energy: 
\begin{equation}
\alpha = \frac{\Delta t F_{\rm per} c_{\rm bol}}{E_b},
\label{eq:alpha}
\end{equation}
where $c_{\rm bol}$ is the bolometric correction giving the inverse fraction of total persistent flux emitted in the instrumental band.
The $\alpha$ value may be calculated using the {\tt alpha} function, given the measurable parameter inputs.

As for the burst rate, we may estimate the average $\alpha$ for a set of $N$ bursts with  
fluences $E_{b,i}$, 
observed in $n$ low-duty cycle observations with persistent flux $F_{{\rm per},j}$ and exposure $T_j$ as
\begin{equation}
\overline{\alpha} = \frac{c_{\rm bol}\sum^n T_jF_{{\rm per},j}}{\sum^N E_b,i}
\label{eq:alphaav}
\end{equation}
assuming a common bolometric correction $c_{\rm bol}$ for each observation.

The $\alpha$ parameter is an important diagnostic of the burst fuel and ignition conditions.
The ``observed'' value given by equations \ref{eq:alpha} and \ref{eq:alphaav} must be interpreted with a little caution, as they likely incorporate the effects of anisotropy of the burst and persistent emission. 
We can also write $\alpha$ as
\begin{equation}
\alpha % & = & \frac{\Delta t F_{\rm per} c_{\rm bol}}{E_B} \\
 =  \frac{Q_{\rm grav}}{Q_{\rm nuc}}\frac{\xi_b}{\xi_p}(1+z).
 \label{eq:alphath}
\end{equation}
Note the implicit dependence of the ``observational'' $\alpha$-value (equations \ref{eq:alpha} and \ref{eq:alphaav}) on both the surface redshift and the anisotropy parameters, illustrated in equation \ref{eq:alphath}. 
The expected value for $Q_{\rm grav}$ is $\approx190$~MeV~nucleon$^{-1}$ (for $M_{\rm NS}=1.4\ M_\odot$ and $R_{\rm NS}=11.2$~km, giving $1+z=1.259$). Thus, excluding the anisotropy effects, $\alpha$ should be $\approx40$ 
for bursts burning H-rich fuel, or 150 
for pure He.

The nuclear energy generation rate is thus
\begin{equation}
Q_{\rm nuc} = \frac{c^2 z}{\alpha}\frac{\xi_b}{\xi_p}. \label{qnuc}
\end{equation}
By substituting the linear expression in equation \ref{eq:qnuc} for simplicity, we  estimate 
\begin{equation}
\Xb
 = z\frac{155}{\alpha}\frac{\xi_b}{\xi_p} - 0.223. \label{xbar}
\end{equation}
This equation is implemented in the {\tt hfrac} function, which takes as input the $\alpha$ value and any inclination constraints.

Clearly this expression is only applicable for $\alpha$-values up to some limit, which we calculate as
\begin{equation}
\alpha \leq 697\, z \frac{\xi_b}{\xi_p}.
\end{equation}
Interestingly, this limit is 
$\approx20$\%
larger than the equivalent value derived for the old expression for $Q_{\rm nuc}$. 
We note that the calculation of \cite{goodwin19a} also took into account the possibility of incomplete burning, which will reduce the $Q_{\rm nuc}$ value.
Nevertheless, this result suggests that larger observed values of $\alpha$ may be accommodated without resorting to explanations including incomplete burning of burst fuel \cite[e.g.][]{bcatalog}.

Conversely, for low values of $\alpha$, we may apply the constraint that $\Xb \lesssim 0.77$, corresponding to the expected maximum possible for accreted fuel with primordial abundances.

The hydrogen fraction at ignition $\Xb$ and the burst recurrence time may then be used to estimate the hydrogen fraction in the accreted fuel, $X_0$. The H-fraction at the base of the fuel layer is reduced steadily by $\beta$-limited hot CNO burning, and will completely exhaust the accreted hydrogen in a time \cite[]{lampe16}
\begin{equation}
t_{\rm CNO} =9.8\left(\frac{X_0}{0.7}\right) \left(\frac{Z_{\rm CNO}}{0.02}\right)^{-1},
\label{eq:tcno}
\end{equation}
measured in the NS frame.
The hydrogen fraction at the base will thus be $X_0[1-\Delta t/(1+z)t_{\rm CNO}]$, 
provided that $\Delta t < (1+z)t_{\rm CNO}$. Once $\Delta t$ exceeds the time to burn all the hydrogen at the base, the abundance there will be zero, and a growing layer of pure He fuel will develop. The average H-fraction in the layer for these two cases will be 
\begin{equation}
\Xb = \begin{cases}
  X_0(1-0.5f_{\rm burn}) & \Delta t \leq (1+z)t_{\rm CNO} \\
  0.5X_0/f_{\rm burn} & \Delta t > (1+z)t_{\rm CNO},
  \end{cases}
  \label{eq:xbar}
\end{equation}
where $f_{\rm burn} = \frac{\Delta t}{(1+z)t_{\rm CNO}}$ is the ratio of the recurrence time to the time to burn all the H.
We can combine this expression with equation \ref{xbar} to give
\begin{equation}
X_0 = \begin{cases}
         z\frac{142}{\alpha}\frac{\xi_b}{\xi_p} - 0.145 
                           +\left[\frac{\Delta t}{(1+z)28\,{\rm hr}}\right]
                       \left(\frac{Z_{\rm CNO}}{0.02}\right) & f_{\rm burn} \leq 1\\
         \sqrt{ \frac{\Delta t}{(1+z)7\,{\rm hr}}\frac{Z_{\rm CNO}}{0.02}
                        \left(z\frac{142}{\alpha}\frac{\xi_b}{\xi_p} - 0.145\right) }
                        & f_{\rm burn} > 1.
        \end{cases}
    \label{eq:X_0}
\end{equation}
Practically, the issue with these expressions is that calculating $f_{\rm burn}$ requires knowledge of $X_0$ (from equation \ref{eq:tcno}), which is of course the unknown we are trying to constrain. We adopt the approach of selecting a trial value of $X_0$, calculating $f_{\rm burn}$ and hence an updated estimate of $X_0$ via equation \ref{eq:X_0}, and iterate until no further change in the estimate arises.
This algorithm is implemented in the {\tt X\_0} function, which is also called by {\tt hfrac} for completeness.

\begin{figure}[ht!]
\epsscale{1.2} % added for preprint version
\plotone{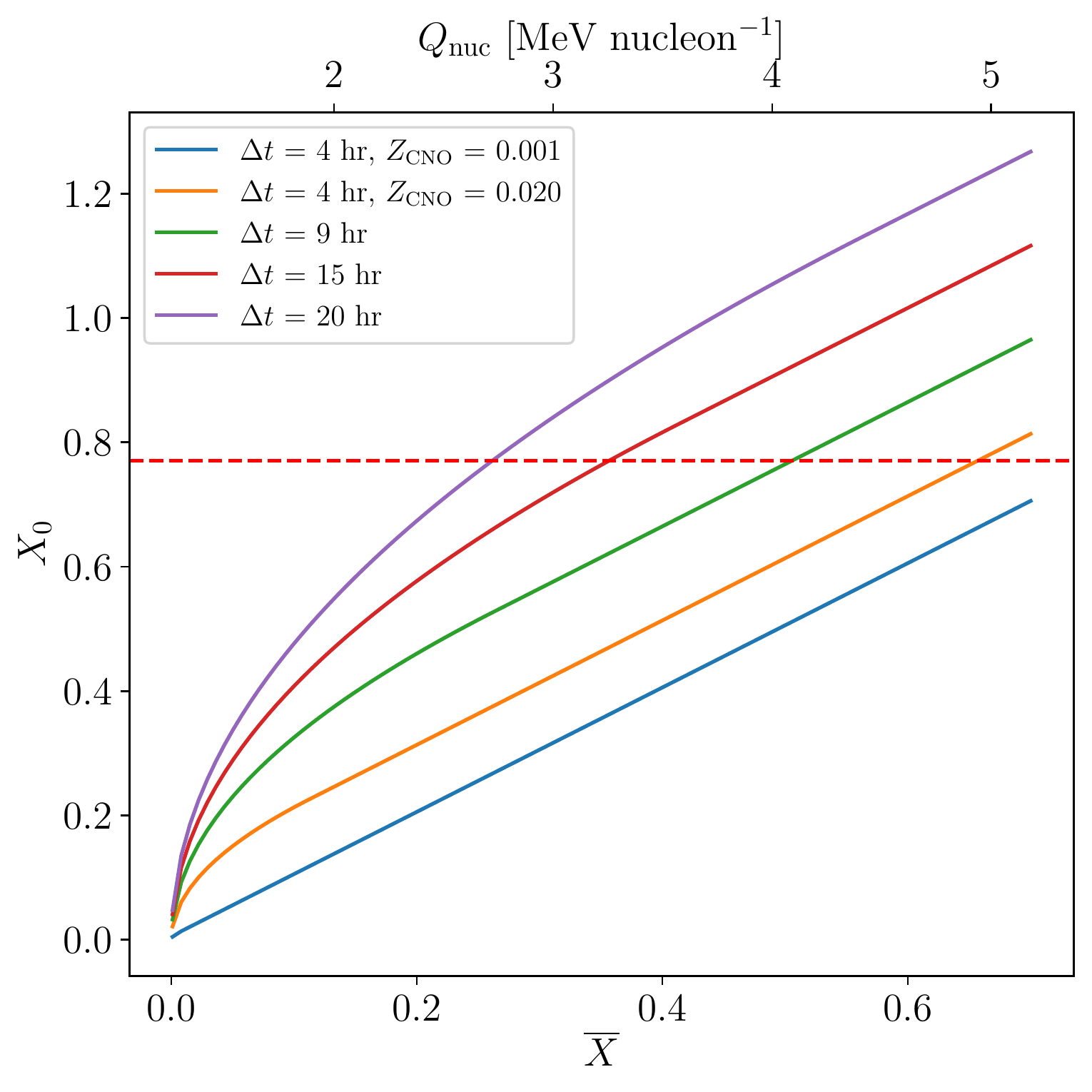}
\caption{Hydrogen mass fraction of the accreted fuel, $X_0$, inferred from the average H-fraction of the burst at ignition, $\overline{X}$, as a function of burst recurrence time $\Delta t$. $X_0$ is calculated from equation \ref{eq:X_0}, assuming $\zcno=0.001$ (blue line) or 0.02 (other lines).
Note how for some combinations of $\overline{X}$, $\Delta t$ the inferred fuel H-fraction is $>0.77$ (red dashed line). Such values are physically implausible, and may be rejected; or alternatively used as evidence for smaller CNO mass fraction $\zcno$   than the adopted value.
The top $x$-axis shows the corresponding $\qnuc$ values.
\label{fig:xbar}}
\end{figure}

We illustrate the relation between $\Xb$ and the derived $X_0$ for various different recurrence times and metallicities in Fig. \ref{fig:xbar}.

Given a minimally complete set of burst observations as described in \S\ref{subsec:data}, a common approach \cite[e.g][]{falanga11} is to estimate the fuel composition at ignition, and hence the accreted fuel composition, based on simple analytic estimates of the burst energy production $Q_{\rm nuc}$.
We explore in the next section how precise such estimates may be, and how the other system parameters can introduce systematic errors.

\subsection{Simulating observations with burst models}
\label{subsec:models}
\label{subsec:simobs}

In order to test the analysis approaches described in this paper, we adopted a set of 60 simulated bursts with the {\sc kepler} code  \cite[]{woos04}, 
as used to measure $\qnuc$ as a function of the input parameters \cite[]{goodwin19a}.
The simulations were carried out on a grid of $\zcno=[0.01, 0.02, 0.1]$ and accretion rate $\dot{m}/\dot{m}_{\rm Edd}=[0.1, 0.2, 0.3]$, where the Eddington accretion rate is defined for the model as
\begin{equation}
    \dot{m}_{\rm Edd}=8.8\times10^4\left(\frac{1.7}{1+X_0}\right)\ {\rm g\,cm^{-2}\,s^{-1}}.
\end{equation}
The simulation results are listed in Table \ref{tab:kepler}.

Each burst train was analysed to identify the bursts (usually discarding a few at the beginning with unusual properties due to insufficient ``burn-in'', 
and calculate the average
recurrence time $\Delta t_{\rm pred}$, burst energy $E_{\rm pred}$ and lightcurve $(t_i,L_{{\rm pred},i})$ covering the extent of the burst.
The model simulates the burning atmosphere in a plane-parallel grid with constant gravity, $g$, an approximation justified by the extreme aspect ratio of the problem.
As a result, the model predictions must be corrected to take into account the general relativistic (GR) effects which are strongest at the NS surface, to derive the quantities that would be measured by an observer.
The input accretion rate $\dot{m}$ may also be converted to a persistent flux level for comparison with observations.

The full set of parameters used to convert the simulations to observed quantities
include the distance $d$; system inclination $i$; and surface redshift $(1+z)$.
We determine the general relativistic (GR) corrections within the constraints of the numerical models, which are typically calculated assuming a Newtonian potential with gravitational acceleration $g=GM/R^2$, where $M$ and $R$ are the equivalent Newtonian mass and radius.
We are free (in principle) to vary $(1+z)$,
for example to achieve improved agreement with an observed recurrence time or burst profile, provided we explicitly maintain consistency with the model $g$ by adjusting the assumed mass and/or radius (see below).

The model lightcurve time bins  $t_i$ and the predicted burst recurrence time $\Delta t_{\rm pred}$ are converted to values as would be measured by a distant observer as follows:
\begin{eqnarray}
t_{i,\infty} & = & (1+z)t_i, \\
\Delta t_{{\rm pred},\infty} & = & (1+z)\Delta t_{\rm pred}.
\end{eqnarray}

Care must be taken to ensure the Newtonian model predictions can be correctly translated to include the GR corrections expected for quantities at the NS surface.
Following \cite{lampe16}, 
we identify a mass and radius for the NS for which the Newtonian acceleration equals the GR value, 
i.e.
\begin{equation}
\frac{GM}{R^2} = \frac{GM_{\rm GR}}{R^2_{\rm GR}\sqrt{1-2GM_{\rm GR}/(c^2R_{\rm GR})}} = \frac{GM_{\rm GR}}{R^2_{\rm GR}}(1+z).
\end{equation}
This equality is generally achieved by assuming that $M=M_{\rm GR}$, and solving for $R_{\rm GR}$. 
We define $\xi$ such that $R_{\rm GR}=\xi R$. 
One advantage of this choice is that the mass accretion rate is identical in the Newtonian and observer frames, and also that $\xi = \sqrt{1 + z}$. In that case, the model-predicted luminosity is related to the luminosity measured by a distant observer, by 
\begin{eqnarray}
L_\infty & = & \frac{\xi^2L}{(1+z)^2} \nonumber \\
& = & \frac{L}{1+z}.
\end{eqnarray}

We note that the combination of model surface gravity $g$ 
and adopted $1+z$  uniquely specifies the NS mass $M_{\rm NS}$ and radius $R_{\rm NS}$:
\begin{eqnarray}
R_{\rm NS,fit} &=& c^2\frac{(1+z)^2-1}{2g(1+z)}, \label{eq:Rimplied} \\
M_{\rm NS,fit} &=& \frac{gR_{\rm NS,fit}^2}{G(1+z)}.
\end{eqnarray}
Thus, by identifying the optimal value of $(1+z)$
for comparison to a particular observation, we can constrain the mass and radius, at a fixed $g$. 
This algorithm  is implemented in the {\tt calc\_mr}  function.

We may also seek to calculate the burst lightcurve that would be observed given a model lightcurve and NS parameters. The process can be summarised as follows. 
\begin{enumerate}
\item Multiply the time bins for the predicted burst lightcurve by the adopted gravitational redshift $(1+z)$, thereby ``stretching'' the profile to account for the general relativistic time dilation at the NS surface
\item Apply the same correction to the model-predicted recurrence time $\Delta t_{\rm pred}$.
\item Interpolate the model-predicted lightcurve onto a set of observational time bins, corresponding (for example) to the typical resolution for time-resolved spectroscopy (0.25~s)
\item Translate the model-predicted luminosity to the corresponding quantity measured by a distant observer, by 
dividing by $(1+z)$
\item Convert the luminosity to (isotropic) flux by dividing by the distance factor, $4\pi d^2$ \label{distfac}
\item Take into account the expected anisotropy effects due to the system inclination, by dividing the luminosity by the anisotropy factor $\xi_b$ \label{anisofac}
\item Calculate the persistent flux expected for the model-assumed accretion rate $\dot{m}$, taking into account (where required) the implied NS radius (equation \ref{eq:Rimplied}) and apply the same corrections as for the burst flux in steps \ref{distfac} and \ref{anisofac} (adopting a separate anisotropy factor $\xi_p$ appropriate for the persistent flux). We also divide by a bolometric correction factor $c_{\rm bol}$ accounting for the limited instrumental passband.
\end{enumerate}
The full procedure is implemented via the {\tt observe} method of  the {\tt KeplerBurst} class.
The simulated lightcurve may then  be compared directly to that observed, either qualitatively, or quantitiatively via a likelihood incorporating the observational errors on each flux bin. The latter approach is implemented via the {\tt compare} method of the {\tt ObservedBurst}  class.

\section{Analysis} 
\label{sec:analysis}

Here we describe different approaches that have been used to deduce system parameters from observations of thermonuclear bursts,
and show how the {\sc concord} code can be used to fully account for the astrophysical uncertainties.
As a companion to this section (and the code itself) we provide a  {\sc jupyter} notebook 
which replicates the analyses described below.

As a simple example of the code utility, we first describe  how the confidence range for peak luminosity may be estimated for a burst from a source with independently known distance. We consider the PRE burst observed from 4U~0513$-$40 with the {\it Rossi X-ray Timing Explorer} ({\it RXTE}) PCA on MJD~54043.68857 \cite[observation ID 92403-01-15-04; MINBAR ID \#3443;][]{minbar}. According to the MINBAR analysis this burst reached a peak flux of $(21.7\pm0.6)\times10^{-9}\ \epcs$. The source is located in the globular cluster NGC~1851, for which the distance is estimated at $(10.32^{+0.20}_{-0.24})$~kpc \cite[]{WatkinsEtAl2015}.

\begin{figure}[ht!]
\epsscale{1.2} % added for preprint version
\plotone{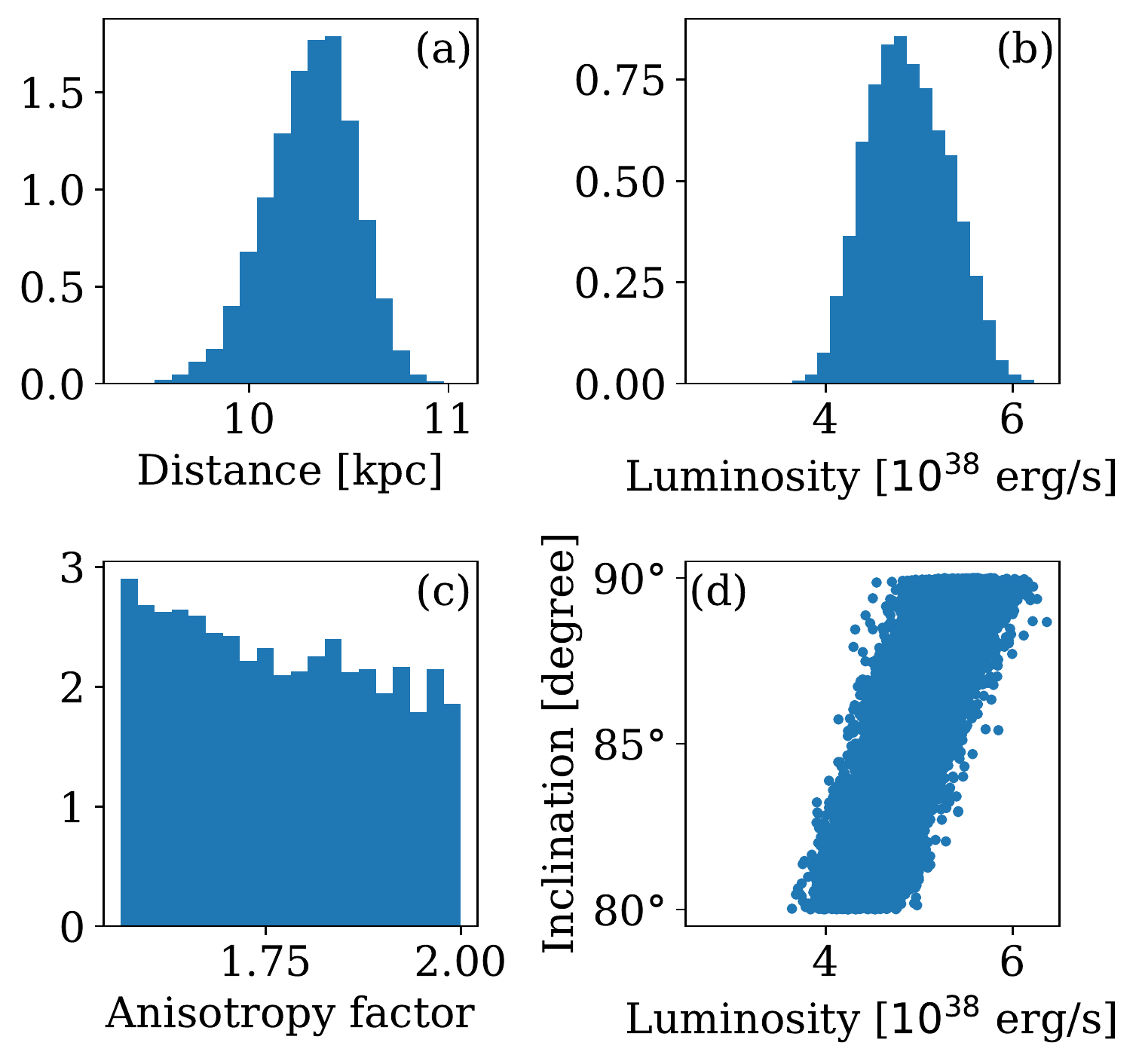}
\caption{Illustration of calculation of burst peak luminosity using {\sc concord}, incorporating measurement uncertainties and the modeled effect of the burst anisotropy. The adopted distance PDF is shown in panel (a); note the slight skew in the distribution. The resulting luminosity PDF in panel (b) is substantially higher than the isotropic value ({\it red dashed line}). The $1\sigma$ confidence intervals are indicated by the green lines. The PDF for the anisotropy factor $\xi_b$ is shown in panel (c); the dependence of the luminosity on the adopted system inclination is shown in panel (d).
\label{fig:burstlum}}
\end{figure}

The isotropic luminosity can be calculated from the peak flux and distance using the {\tt luminosity} method, with {\tt isotropic=True} as $2.77\times10^{38}\ \eps$. 
However, a more accurate calculation should take into account the uncertainty in the flux; the (asymmetric) uncertainty in the distance; and the likely effect of the anisotropy in the burst emission, based on the estimated system inclination $i>80^\circ$ \cite[]{fiocchi11}. This can be achieved by \edit1{pre-calculating the distance {\sc astropy} distribution (including the anisotropic errors, and the correction required for cumulative statistics; Figure \ref{fig:burstlum}, panel a), and passing this distribution as well as the flux uncertainty to the routine; a (symmetric normal) distribution is also generated for the flux}, with the required properties. We also pass the inclination limits, and an array of inclination values is generated assuming an isotropic distribution of inclinations within the limits (i.e.  uniform in $\cos i$).
The burst anisotropy factor is calculated automatically for each inclination value, based on the default model.
The resulting luminosity is significantly higher, 
$(4.86_{-0.42}^{+0.48})\times10^{38}\ \eps$
(Fig. \ref{fig:burstlum}, panel b); the discrepancy is primarily the result of the significant attenuation of the burst flux from the viewing angle close to the plane of the accretion disk. The corresponding anisotropy value $\xi_b$ is in the range 1.56--2 (Fig. \ref{fig:burstlum}, panel c). The dependence of the luminosity on the assumed inclination is illustrated in Fig. \ref{fig:burstlum}, panel d. 

The output from the function is by default the median value and the lower and upper uncertainties, at the required confidence level (68\% by default). Using the {\tt fulldist=True} flag, the function instead returns a dictionary including the full distributions for the luminosity, distance, inclination, anisotropy factor, as well as the anisotropy model identifier. % and the confidence level
Each of the distribution arrays can be accessed via the {\tt distribution} attribute.
This object can be written to a file or used for input to subsequent analyses.

In the sections below we describe other examples,
focussing on adopting approaches tailored to the availability of data, which will differ for different sources. 

\subsection{Code validation}
\label{subsec:validation}

We next seek to validate the approach adopted to measure the fuel composition in \S\ref{subsec:energy}, by comparing inferred quantities to independently-measured values. We make use of the set of {\sc kepler} simulations 
described in \S\ref{subsec:models},
which were carried out over a range of accretion rates and compositions ($X_0$, $\zcno$) to measure the $\qnuc$ value. For each of 60 simulations, \cite{goodwin19a} measured the average burst energy and recurrence time (in the model Newtonian frame) and the average hydrogen fraction of the fuel layer, $\xbar$.  

\begin{figure}[ht!]
\epsscale{1.2} % added for preprint version
\plotone{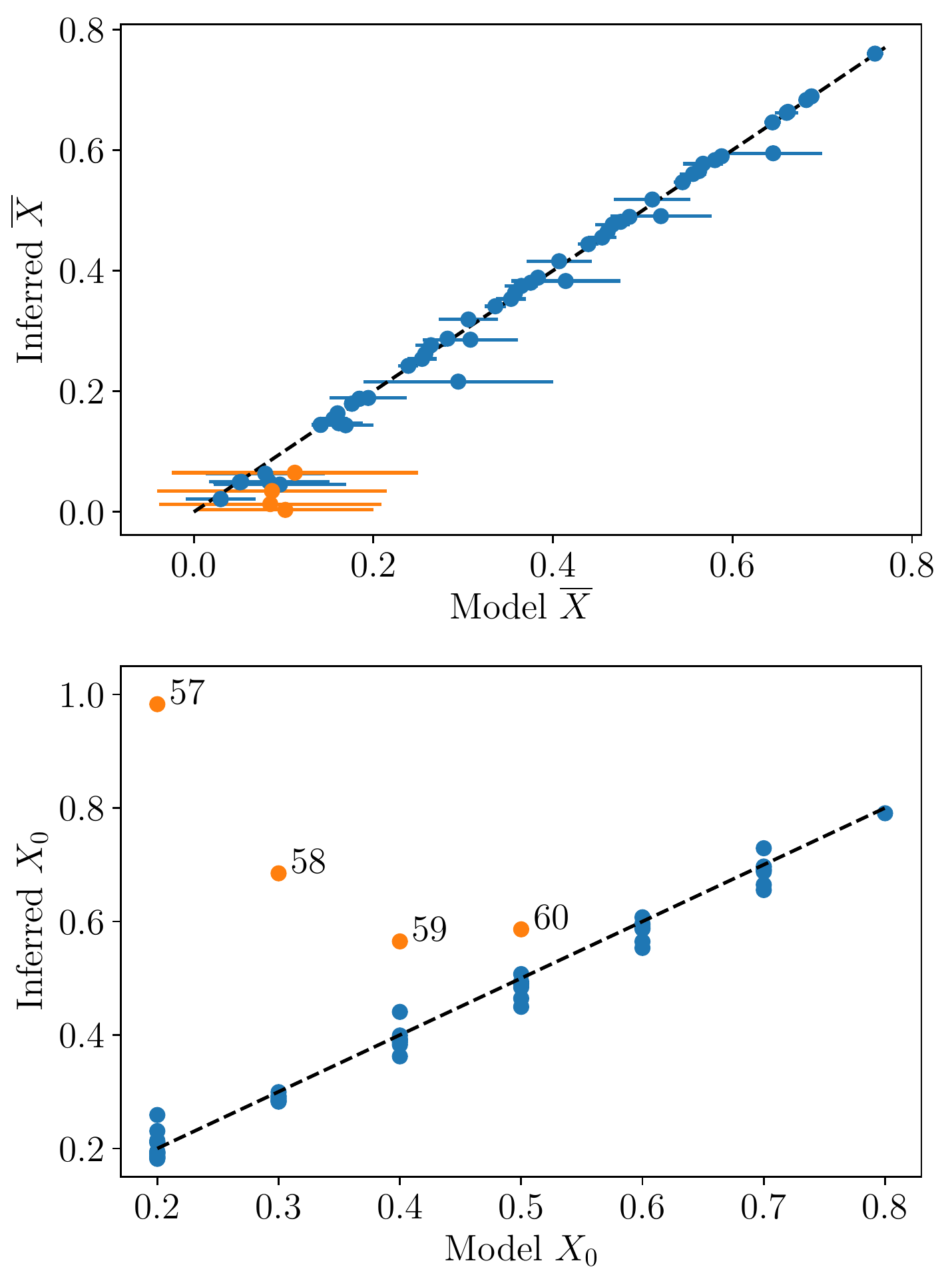}
\caption{Validation of the $\xbar$ and $X_0$ calculations (equations \ref{eq:tcno}--\ref{eq:X_0}) by comparison with {\sc kepler} model runs over a range of compositions and accretion rates. The top panel shows the inferred average hydrogen mass fraction of the burst at ignition, $\overline{X}$, as a function of the value measured from the model. The RMS error is 0.025, but the burst-to-burst variation is significantly larger for some of the runs reaching low $\Xb$ (illustrated by the horizontal error bars).
The highlighted points are the outliers identified from the lower panel, which shows the inferred H-fraction of the accreted fuel, $X_0$, inferred from the recurrence time and input composition.
Excluding the highlighted and labeled points, the calculation is reasonably accurate, with an RMS error of 0.021. 
\label{fig:validate}}
\end{figure}

Based on the input composition and the burst recurrence time, we first calculated $\xbar$ from equations \ref{eq:tcno} and \ref{eq:xbar}, and compared to the value measured from the model (Fig. \ref{fig:validate}, top panel). We find generally a good agreement, with overall RMS error 0.025, but significantly larger fractional errors at low $\Xb$. We also calculated $X_0$ from equation \ref{eq:X_0}, which is shown against the model input values in Fig. \ref{fig:validate}, bottom panel. For most of the runs the accuracy is reasonable, with RMS 0.021, but there are some notable outliers with much larger errors. These runs (\#57--60) are associated with high $\zcno=0.1$ and significant variations in $\Xb$ from burst-to-burst. 

\begin{figure}[ht!]
\epsscale{1.2} % added for preprint version
\plotone{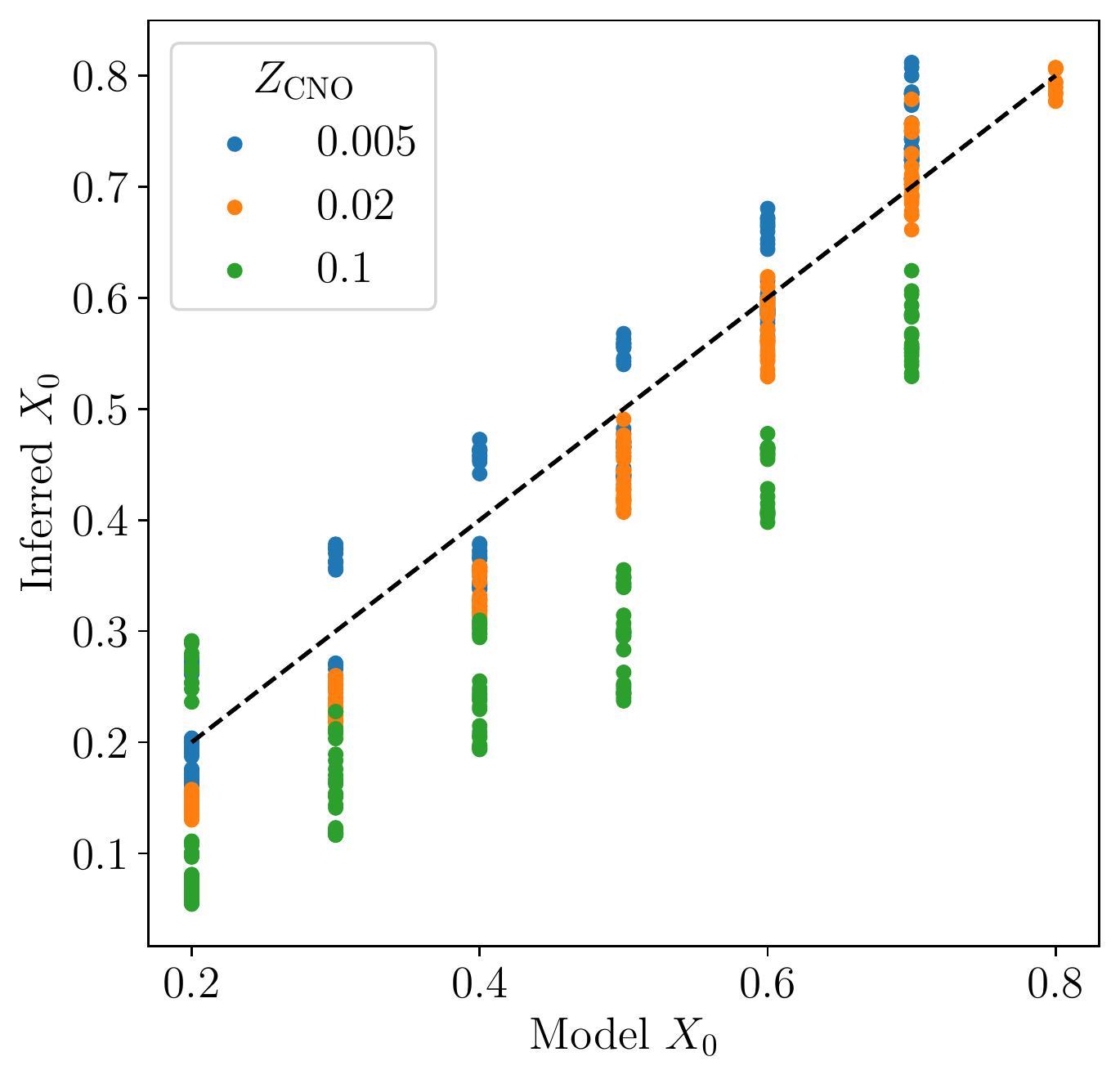}
\caption{Inferred $X_0$ based on simulated observations from the {\sc kepler} model run set, with randomly-distributed inclinations in the range 0--$75^\circ$. The symbol colour shows the 
metallicity $\zcno$ for the simulation.
The 
NS redshift and metallicity $\zcno$ are 
left at the {\tt concord} defaults for the calculation. Note how the deviation from the 1:1 line is greatest where the model input metallicity is furthest away from the assumed value, i.e. 0.1 ({\it green symbols}), compared to the default value of 0.02.
\label{fig:X0}}
\end{figure}

Each of these tests so far are somewhat unrealistic, as they do not include the confounding effects of anisotropic emission and other factors. Thus, we also generated simulated observations to infer $X_0$, following the approach in \S\ref{subsec:fuelcomp}. We transformed the model-predicted values to observational quantites at a fixed distance $d=6$~kpc, NS redshift $1+z=1.259$, and metallicity $\zcno=0.02$. We adopted 10 different isotropically-distributed inclinations in the range 0--$75^\circ$ for each model run, giving a total of 600 simulated data sets. We then inferred the hydrogen fuel fraction $X_0$ using the {\tt hfrac} method but assuming no knowledge about the inclination. For each instance a range of possible values of $X_0$ is obtained, based on the uncertainties in the input parameters and the possible range for the inclination $i$. The anisotropy factors were calculated using ``model A'' of \cite{he16}, which is incorporated into the {\sc concord} suite. We show the median inferred H-fraction values as a function of the model input values in Fig. \ref{fig:X0}. There is a moderately large scatter about the 1:1 line, with RMS error 0.097, which obviously will have a more substantial fractional effect at the low-$X_0$ end of the range.

We understand the scatter in Fig. \ref{fig:X0} as arising from a mismatch in parameter space, i.e. where the assumed parameters are different from the model input parameters. The choice to estimate $X_0$ over a distribution of inclination values is intended to address this mismatch, at least in an average sense. However by default the {\tt hfrac} method uses a fixed $\zcno=0.02$, while the model runs include both larger and smaller values, 0.1 and 0.005 respectively. We find that the largest errors in the measured $X_0$ come from the runs with $\zcno=0.1$; if we exclude these runs from our simulations, we find an RMS error just over half as large as for the full sample, at 0.051. Clearly the typical mismatch error will depend on the underlying distribution of $\zcno$ (i.e. the prior), and so this distribution is important to consider for optimal precision \cite[cf. with][]{goodwin19c}

Introducing a range for the NS redshift to these simulations is constrained by the fixed (Newtonian) surface gravity $g$ chosen for the simulations. Nevertheless, this would be possible, along with allowing a distribution for the metallicity $\zcno$. It is to be expected that the error would increase with additional parameter freedom, although to fully quantify this effect would also need suitable priors for those parameters. On the other hand, some of the degeneracy may be resolved by simultaneously analysing pairs of bursts measured at different accretion rates. We now apply the {\sc concord} tools to further examples of observational data.

\subsection{Two or more regular, consistent bursts} 
\label{subsec:fuelcomp}

The ideal observational situation is when we have two (or more) bursts detected by sensitive X-ray instruments (allowing precise measurements of flux and fluence), and can confidently infer the recurrence time, 
either because the bursts are observed in observations without data gaps, or where the bursts are sufficiently regular that the recurrence time can be confidently constrained {\it despite} any gaps.
Here we require measurements of the burst recurrence time $\Delta t$, fluence $E_b$, and the persistent flux at the time of the bursts, $F_{\rm per}$. 
We caution that it is likely important to ensure that the burst observations comprise an extended sequence of more than just a few events, due to the episodic nature of some burst behaviour \cite[e.g. short recurrence time bursts;][]{keek10}.
We use here as an example, the three trains of bursts observed with the \xte/PCA from GS~1826$-$238, as analysed by \cite{gal17a}. 

We use the {\tt alpha} function to first determine the ratio of burst to persistent luminosity, taking into account the uncertainties on each of the measurables. Applying this to the burst train observed in 
1998 June, with  $\Delta t=(5.14\pm0.07)$~hr, 
we find 
$\alpha=35.4\pm0.7$,
which is at the lower limit of the expected range of values, 
consistent with the expected 
H-rich fuel in this system (cf. with \citealt{johnston20}).
However, this value 
includes 
the possible effects of system anisotropy, for the inferred range of system inclinations of 
$i = 69^{+2}_{-3}$~degrees.

Thus, we can go further by estimating the fuel composition at ignition, and as accreted, using the {\tt xbar} routine. We adopt an initial isotropic distribution of inclinations within the above range. 
We find inferred values of $X_0$ in the range 0.5--0.7, only minimally overlapping with the range inferred by \cite{johnston20}. The discrepancy becomes worse if we consider the other two epochs of bursts from \cite{gal17a}, each of which has an even lower inferred range of $X_0$.

\begin{figure}[ht!]
\epsscale{1.2} % added for preprint version
\plotone{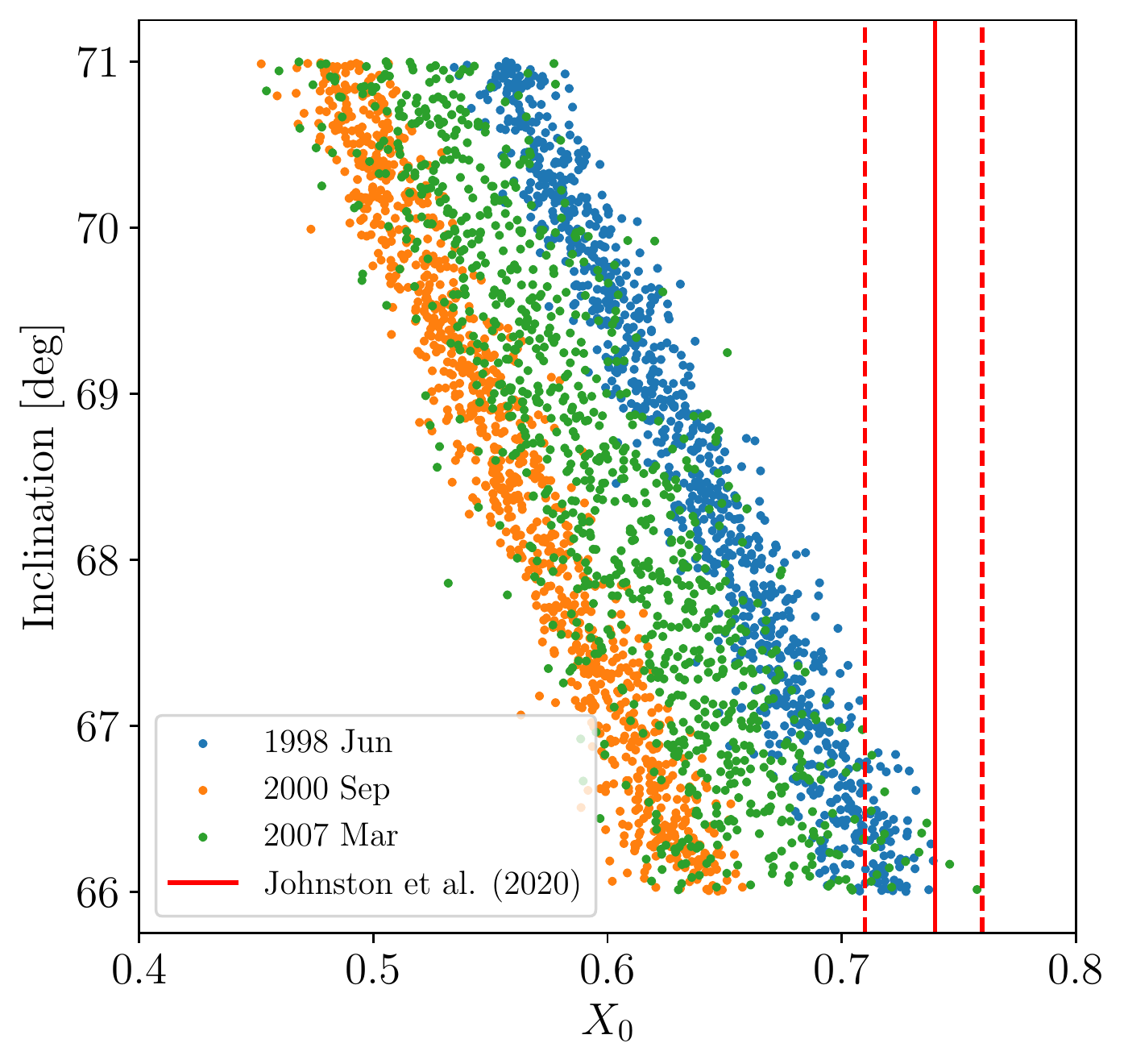}
\caption{Constraints on the fuel composition for the ``Clocked burster'', GS~1826$-$238, based on the pairs of bursts observed with \xte/PCA analysed by \cite[]{gal17a}. Histograms of the $X_0$ values might lead to the conclusion that the overlapping region could be taken as the best estimate of the PDF for the combined dataset. However, plotting this parameter as a function of the system inclination $i$ demonstrates the incompatibility of the three datasets.
\label{fig:X_0_1826-238}}
\end{figure}

These constraints are derived by adopting a fixed value of $\zcno=0.02$; we can also adopt a distribution of values instead (effectively as a Bayesian prior), but this approach will only broaden the resulting histograms, not addressing the systematic discrepancy.
The tension between these three epochs is illustrated in Figure \ref{fig:X_0_1826-238}, which shows each epoch defining a partially distinct region in $X_0$--$i$ parameter space.
In particular the 1998~June and 2000~September epochs appear to offer no overlap in parameter space, suggesting their properties cannot be reconciled with a single set of system parameters.
Although this result might appear to be a failure of the calculations, we examine the detailed implications further in the discussion.

\subsection{One or more bursts, no recurrence times} 
\label{subsec:xmmu}

The situation may arise where a number of bursts $N$ have been observed in low-duty cycle observations, such that no pair of bursts are sufficiently close together to really constrain the recurrence time. 
\edit1{
The average burst rate $\overline{R}$ can be estimated based on the total exposure, i.e.
\begin{equation}
\overline{R} = \frac{N}{\sum^n T_i},
\label{eq:rate}
\end{equation}
where each observation $i$ has exposure $T_i$. 
The probable range of the average rate can be estimated assuming Poisson counting statistics for the uncertainty on the burst number $N$. This approach is implemented in the code as the {\tt tdel\_dist} function, which takes as arguments the number of bursts detected, and the total exposure.
Because of the approximately periodic behavior common for bursts, and the similarity between low-Earth orbital periods and the burst recurrence times, such uncertainty estimates may be wildly incorrect.}

We choose as an example the \xte/PCA observations of the bursting source XMMU~J181227.8$–$181234 analysed by \cite{goodwin19b}, which included 7 bursts detected within a set of observations with total exposure 0.3446~d. Since one of these bursts was interpreted as a secondary event of a short-recurrence time multiple, the analysis adopted $N=6$. We estimate the PDF for the recurrence time using the {\tt tdel\_dist} function, by estimating the PDF for the underlying rate, such that 6 bursts would be observed. We can then calculate the confidence intervals for the rate. %, but we can also go further.

Based on the average burst fluence of 
$(24\pm7)\times10^{-9}\ \epc$ and the average persistent flux of $(1.19\pm0.16)\times10^{-9}\ \epcs$, we can use the inferred distribution of recurrence times to calculate a corresponding distribution of $\alpha$, using the {\tt alpha} function. The resulting distribution has very high values, with a 68\% range of 360--1000. We can then use the {\tt hfrac} function to determine the H-fraction at ignition $\overline{X}$, and of the accreted fuel $X_0$, which also depends on assumptions about the (unknown) system inclination.
The functional form of the $X_0$ dependence on the other parameters (equation \ref{eq:X_0}) means that, for high $\alpha$-values, we tend to derive {\it negative} $X_0$ values. Clearly, these can be rejected as unphysical, and in fact for this source, only a small fraction ($\approx2$\%) of the derived values will result in $X_0>0$.

\begin{figure}[ht!]
\epsscale{1.2} % added for preprint version
\plotone{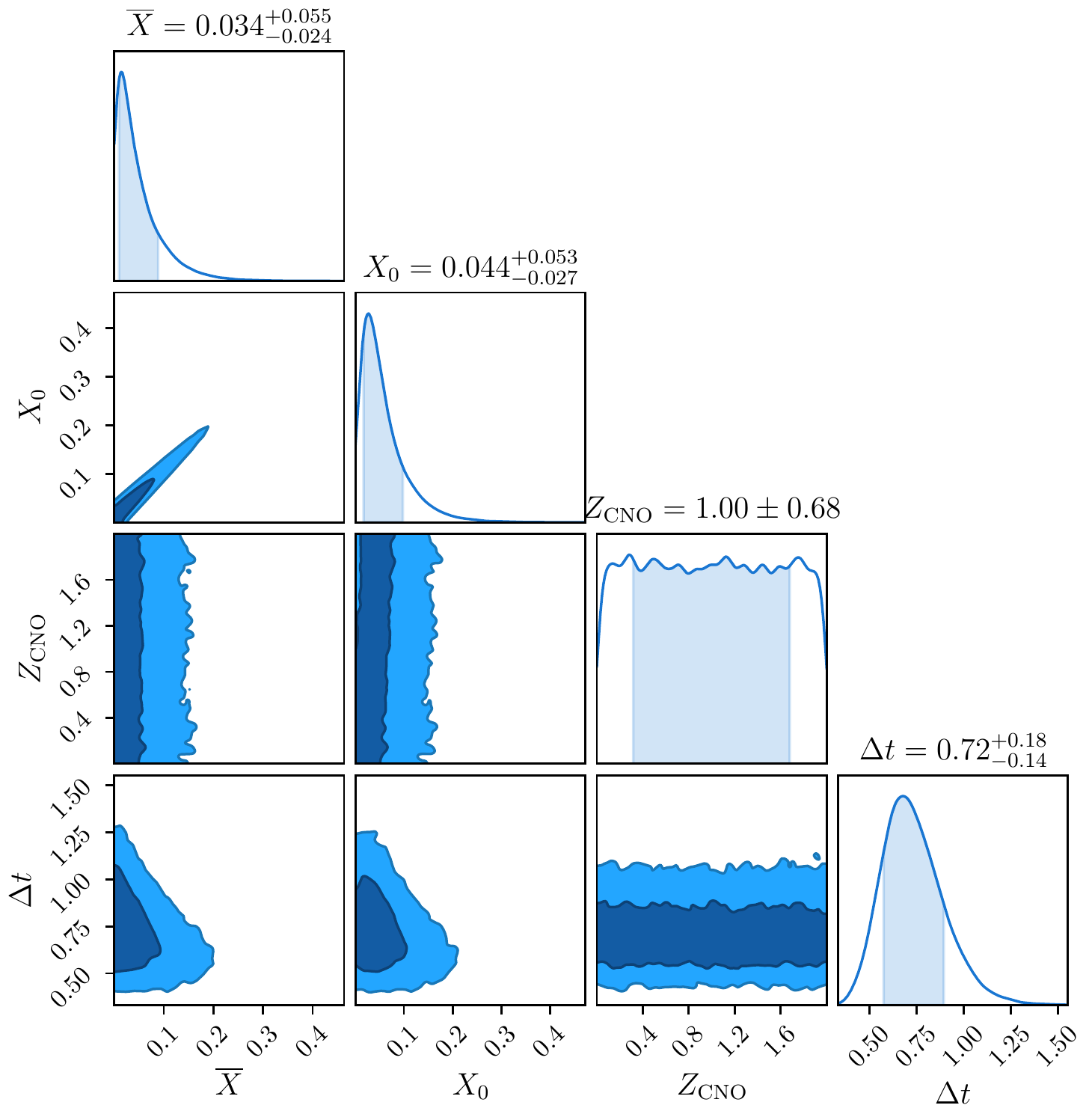}
\caption{Inferred distributions of burst parameters for 6 events observed from XMMU~J181227.8$-$181234 with \xte/PCA, as reported by \cite{goodwin19b}. The number of bursts and the exposure are used to estimate the PDF of the recurrence time $\Delta t$, and with the persistent flux, the $\alpha$-values. The burst fluence is then used to estimate the  H-fraction at ignition (and accreted), $\overline{X}$ and $X_0$ respectively, and we select only the physically realistic values $X_0>0$. The resulting constraint on the fuel H-fraction $X_0$ is extremely strong, but for $\zcno$ \edit1{(here given as a percentage)} less so as the burst fuel is intrinsically so low in H, and the recurrence time (here in units of hr) is short.
\label{fig:param_xmmu181227}}
\end{figure}

The inferred \edit1{hydrogen} fraction at ignition is very low (Fig. \ref{fig:param_xmmu181227}), which, coupled with the short recurrence times, infers a similarly low value for the accreted \edit1{fraction} ($X_0<0.15$ at 95\% confidence). These constraints are comparable to those reported by \cite{goodwin19b}, although slighly broader for $\overline{X}$, $X_0$ as a distribution in $\zcno$ is allowed, while narrower in the recurrence time $\Delta t$ as the self-consistency selection on $X_0$ is also applied to the inferred distribution of $\Delta t$.

\subsection{One burst -- estimating the distance and burst rate} 
\label{subsec:1burst}

For a single burst detected from a bursting source, provided the data quality and signal-to-noise are sufficient, we can expect to measure the burst peak flux, fluence, and timescale. If the burst exhibits PRE, we can infer the distance (\S\ref{subsec:distance}); from the persistent flux, the accretion rate; and (via burst models) the expected burst rate. The detailed shape of the lightcurve can provide constraints from comparisons with numerical models (see \S\ref{subsec:lccompare}), but here we focus on the more straightforward calculation which may be done from the simple burst measurements alone.

We use as an example 
the event detected with \igr/JEM-X from IGR~J17591$-$2342 \cite[]{kuiper20}. This burst was observed  when the estimated persistent bolometric flux was $(1.2\pm0.2)\times10^{-9}\ \epcs$, and exhibited a fluence of $E_b =(1.1\pm0.1)\times10^{-6}\ {\rm erg\,cm^{-2}}$. The peak flux was $(7.6\pm1.4)\times10^{-6}\ \epcs$.

\begin{figure}[ht!]
\epsscale{1.2} % added for preprint version
\plotone{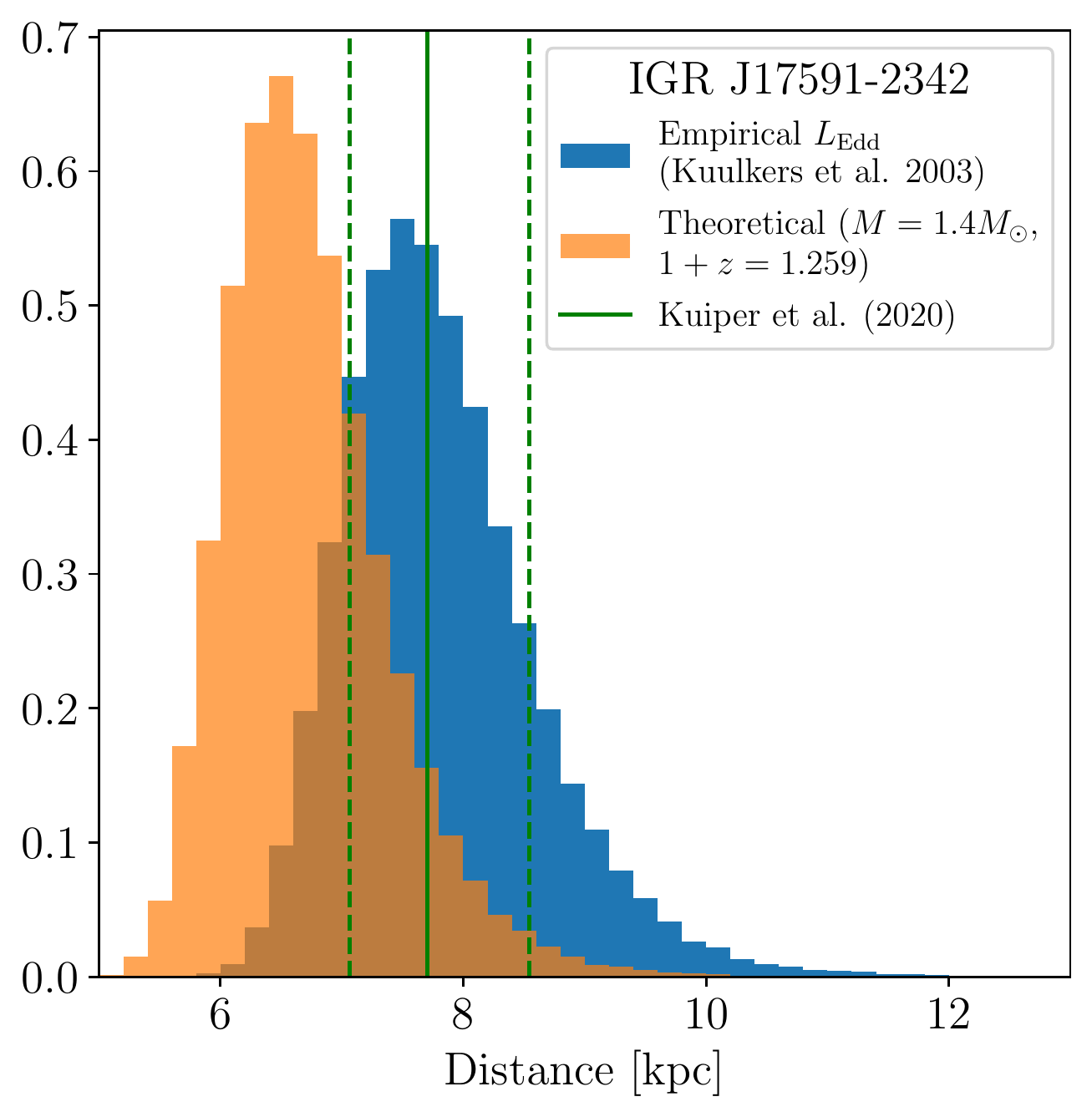}
\caption{Distance estimates for IGR J17591$-$2342, based on the single PRE burst detected with JEM-X \cite[]{kuiper20}. The blue histogram and green 1-$\sigma$ limits replicate the analysis in that paper, comparing the flux to the empirical Eddington limit of \cite{kuul03a}, and adopting the inclination limis of \cite[]{sanna18}.  For comparison we also include the inferred distance distribution (orange histogram) adopting the theoretical Eddington luminosity, for a NS mass of $1.4M_\odot$, and surface redshift of $1+z=1.259$.
\label{fig:dist_igr17591}}
\end{figure}

Following \cite{kuiper20}, we first estimate the distance to the source based on the empirical value of the Eddington luminosity (using the {\tt dist} method with {\tt empirical=True}), and incorporating the expected anisotropy for an inclination range of 
24--30$^\circ$ \cite[]{sanna18} as $7.7_{-0.6}^{+0.8}$~kpc. Here the uncertainty is dominated by the peak flux, since the anisotropy factor is effectively fixed for such a narrow range of inclination, at $\xi_b\approx0.70$.
The PDF for the distance, which is distinctly non-Gaussian, is plotted in Figure \ref{fig:dist_igr17591}. We also show the corresponding distribution resulting from adoption of the theoretical Eddington luminosity, according to equation \ref{ledd}.

The {\sc concord} code allows us to pass the derived distance distribution directly to the method for calculating the accretion rate at the time of the burst, also incorporating the dependence of the persistent flux anisotropy (which is not the same as for the burst flux).
The persistent flux can then be converted to an accretion rate, given the estimated anisotropy factor predicted (separately) for the persistent emission, 
by inverting equation \ref{eq:fper}, via the {\tt mdot} method, and incorporating the estimated distance distribution already determined. 
Based on the distance constraints, the estimated accretion rate  is within the range $(2.0_{-0.4}^{+0.6})\times10^3\ \rm{g\,cm^{-2}\,s^{-1}}$ for the usual assumed values for NS mass and radius (and hence redshift).

\begin{figure}[ht!]
\epsscale{1.2} % added for preprint version
\plotone{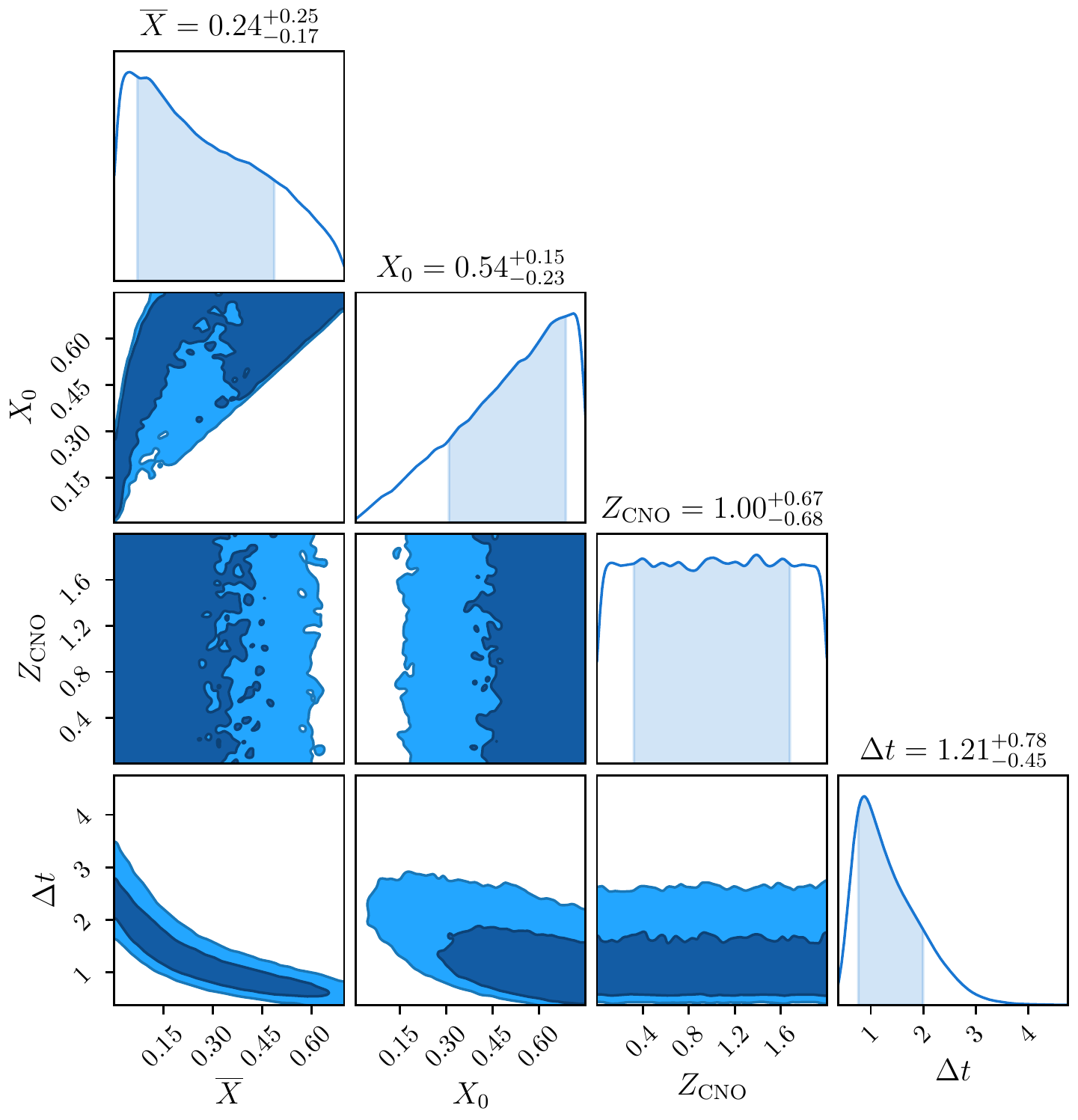}
\caption{Inferred distributions of burst parameters for the single event observed from IGR~J17591$-$2342 with \igr/JEM-X, as reported by \cite{kuiper20}. An initially uniform distribution of \Xb, the mean H fraction at ignition, is used to estimate the burst ignition column $\yign$\ from the measured fluence, $E_b$. The column $\yign$\ and accretion rate are then used to estimate the recurrence time $\Delta t$ and the accreted H-fraction $X_0$, and we select only the physically realistic values $X_0<0.75$. The resulting constraints on the fuel composition parameters $X_0$ and $\zcno$ \edit1{(here given as a percentage)} are rather weak, but the recurrence time $\Delta t$ 
(here in units of days) % originally 
is somewhat better constrained.
\label{fig:param_igr17591}}
\end{figure}

Given a numerical burst ignition model, we could now go on to estimate the recurrence time, and check if it was consistent with the detection of only a single burst over the entire observed outburst. However, there is an extremely wide range of burst recurrence times at fixed accretion rate, depending upon the unknown fuel composition. We could potentially obtain constraints on the fuel composition by comparison of the observed lightcurve to simulations, but it's also possible that multiple different combinations of accretion rate and composition could give rise to similar composition at ignition (and hence shape of the lightcurve).

Instead, we can constrain the fuel composition via the ignition column (equation \ref{eq:column}), based on an estimate of the nuclear burning yield, $\qnuc$. Since this quantity is unknown, we assume a set of random uniformly-distributed\footnote{Such distribution may not be realistic, but it serves to demonstrate the code capabilities. The largest (and \edit1{a priori} most unlikely) values of \Xb\ are suppressed by the limit on $X_0$ imposed during this process.} values of \Xb\ in the range 0--0.7
to estimate $\yign$ from equation \ref{eq:column}.
We then estimate the recurrence time for each pair of $\yign$, $\dot{m}$ values as
\begin{equation}
\Delta t_{\rm rec} = (1+z)\yign/\dot{m},
\end{equation}
and finally calculate the fuel composition, $X_0$, for each set of \Xb, $\yign$, $\Delta t$ from equation \ref{eq:X_0}, again assuming a uniform distribution for $Z_{\rm CNO}$ between 0--0.02.
Because we have chosen uninformative prior distributions for \Xb\ and $Z_{\rm CNO}$, some of the $X_0$ values are unrealistic, well in excess of the maximum expected for such sources, 0.75 or so. We thus limit the distributions of each of the parameters to those such that $X_0\leq0.75$, and quote the $1\sigma$ parameter ranges for each.

The inferred ignition column is $1.63_{-0.58}^{+1.05}\times10^8\,{\rm g\,cm^{-2}}$, and the average expected recurrence time is 
$1.21_{-0.45}^{+0.78}$~d.
We can also infer lower (upper) limits on $X_0$ ($Z_{\rm CNO}$), although these limits are not strongly constraining; we find $X_0>0.17$ and $Z_{\rm  CNO}<0.017$ at 95\% confidence.
The resulting constraints are a marked contrast to those of 
GS~1826$-$238 (Fig. \ref{fig:X_0_1826-238})
and XMMU~J181227.8$-$181234 (Fig. \ref{fig:param_xmmu181227}),
and are a testament to the ability to discriminate between different accreted compositions of these methods, even when the available data is limited.

\subsection{Zero bursts -- constraining the distance}
\label{subsec:zerobursts}

While it may seem strange to base an analysis on the {\it non-}detection of bursts, practically even for the best-studied sources the X-ray observation duty cycle is of order a few percent \cite[e.g.][]{minbar}. There is a reasonable probability that {\it every} burst will be missed in a series of observations, particularly if the accretion rate is low (and hence the bursts are infrequent). In extreme cases this can mean that no bursts whatsoever are observed,  which may also be explained by the compact object being a black hole rather than a NS.

In cases where we can be confident that the compact object is a NS (e.g. where persistent pulsations are detected) but no bursts are observed, we can constrain the distance by adopting a composition for the fuel and comparing the predictions of burst models to the good-time intervals of our X-ray data.

For example, based on {\it RXTE}/PCA observations of the accretion-powered millisecond pulsar IGR~J00291+5934, \cite{gal06b} derived joint constraints over the distance $d$ and fuel H-fraction $X_0$, adopting  the predictions of a simple numerical model 
\cite[see also][]{goodwin19c}. The method is relatively simple\footnote{Although the method is straighforward, we choose not to implement it in {\sc concord}, as this is a relatively niche case. We include it as an example here for completeness.}; 
the persistent flux history is converted to an accretion rate history given an assumed distance (via equation \ref{eq:fper}), and a burst train is generated. The times of the predicted bursts are compared to the good time intervals of the observations, and the simulation is repeated over a range of distances and times for the first burst. The likelihood for a given distance is then estimated as the fraction of simulated burst trains for which all the bursts fall within data gaps, and hence are missed. 

Larger distance (or H-fraction) will imply larger burst rates, and hence a smaller likelihood of missing any. The constraints obtained suggest the distance is $\lesssim6$~kpc (at 3-sigma significance), provided $X_0\approx0.7$, a reasonable choice given the expectation of a H-rich donor in this $2.46$~hr binary orbital period system \cite[]{gal05a}.  
A separate argument is used to give a lower limit on the distance.

\begin{figure}[ht!]
\epsscale{1.2} % added for preprint version
\plotone{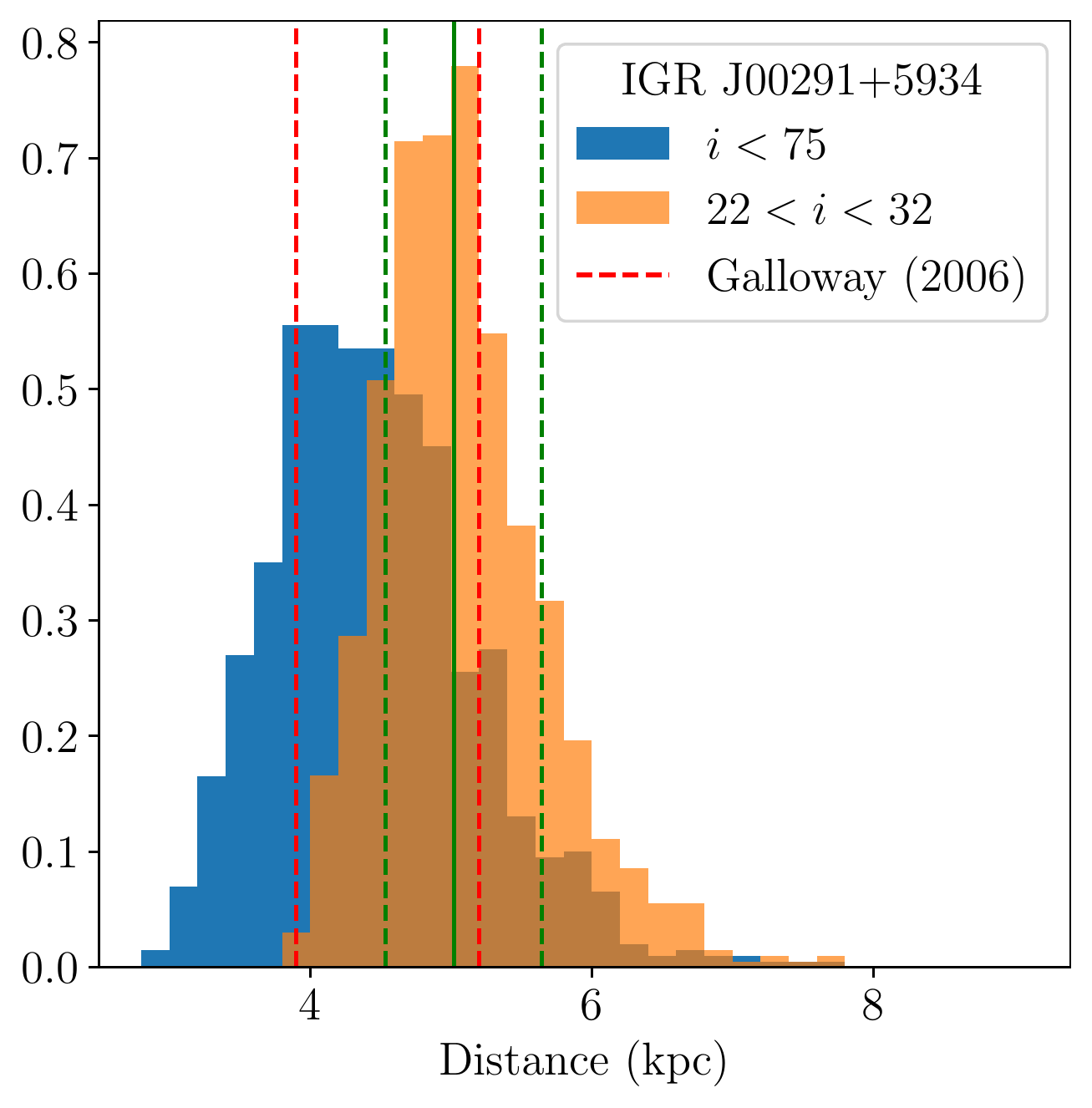}
\caption{Estimated distance for the accretion-powered X-ray pulsar IGR~J00291+5934, based on the peak flux of the sole thermonuclear burst observed by \citet[][solid histograms]{defalco17};
and based on the {\it non-}detection of bursts by \citet[][red dashed lines]{gal06b}.
Note the good agreement between the methods.
The distribution for an isotropic distribution of system inclinations (up to a maximum value of $75^\circ$, motivated by the lack of dips in the X-ray intensity) is shown as the blue shaded histogram. Imposing the inclination constraint of 22--32$^\circ$ suggested by \cite{torres08}, for which the predicted burst anisotropy factor is $\xi_b=0.704$ on average, gives instead the orange-shaded histogram. The resulting $1\sigma$ confidence interval (green lines) is $5.0_{-0.5}^{+0.7}$~kpc.
\label{fig:dist_igr00291}}
\end{figure}

In a subsequent outburst in 2015, a single burst was detected by {\it Swift}/XRT \cite[]{kuin15}, offering the opportunity to verify the previously-determined distance range. The estimated peak 
0.1--35~keV flux was $(18\pm4)\times10^{-8}\ \epcs$ \cite[]{defalco17}, with the burst exhibiting spectral variations indicative of PRE. By comparing the peak flux to the empirical Eddington luminosity (see \S\ref{subsec:1burst}), the corresponding distance 
\cite[including the inclination range of 22--32$^\circ$ suggested by][]{torres08},
is $5.0_{-0.5}^{+0.7}$~kpc, which is fully consistent with the previously established limit. The effect of the inclination constraints is illustrated in Fig. \ref{fig:dist_igr00291}.

\subsection{Observation-model comparisons} 
\label{subsec:lccompare}

We now consider application to comparisons of the measured quantities to the predictions of numerical models. In the simplest case, the model will make predictions for the burst recurrence time $\Delta t$ and fluence $E_b$ given an accretion rate and composition.
More detailed time-dependent models may also provide a burst lightcurve, which can also be compared to the observed lightcurves.
In both cases, the model-predicted quantities must then be converted to what a distant observer would see, by taking into account the effects of source distance, emission anisotropy, general relativistic time dilation and redshift.

We revisit the example of 4U~1820$-$30 described in \S\ref{subsec:fuelcomp}. For that system, we infer $X_0=0.17$ for $\zcno=0.02$, giving a recurrence time (at the accretion rate observed on 1997 May 4 of $0.144\ \dot{m}_{\rm Edd}$) of $\Delta t=2.681$~hr. 
We use the {\sc settle} code 
\cite[]{cb00,cumming03}
to predict the expected conditions for the bursts with these input parameters. Some care must be taken to ensure that the accretion rate is defined according to a consistent value of $\dot{m}_{\rm Edd}$, and also in a consistent reference frame.
The model predicts a recurrence time of 2.24~hr, and burst energy of 
$3.13\times10^{39}$~erg,
and with a model-predicted $\alpha=161$.
Now the predicted burst parameters are already redshifted to the values that would be inferred by a distant observer, based on the NS mass and radius provided to the code.
The comparison with the recurrence time is already fairly reasonable, but we can also compare the other measurable parameters with the model inputs and/or predictions.
We adopt the distance for the source of 7.6~kpc \cite[]{gal17a}, and a system inclination of $i=50^\circ$ for illustrative purposes.

Using the {\tt fper} method, which implements equation \ref{eq:fper}, we can calculate the equivalent
persistent flux (incorporating the bolometric correction)
as $3.26\times10^{-9}\ \epcs$. Using the {\tt lum\_to\_flux} function, and neglecting any bolometric correction (since the measured fluence is already bolometric), we predict a burst fluence of $0.505\times10^{-6}\ \epc$. Each of these predictions are within a few tens of percent of the observed values, but may also be fine-tuned by judicious choice of system parameters $\mns$, $\rns$, $d$, $i$, $X_0$ and $\zcno$. % and Qb

We can go a step further with our model-observation comparisons and incorporate the calculation above into an MCMC code to constrain the system parameters, as implemented by {\sc beans} \cite[]{goodwin19c}. For each set of system parameters, we run the burst model for a burst train, or as many separate burst measurements as are provided. We transform the model predictions to simulate observations, taking into account the system distance and inclination via the anisotropy model, and calculate a likelihood based on the comparison of the model-predicted values and the observations:
\begin{eqnarray}
\mathcal{L} & = & -\left[\left( \frac{F_p-F_{p,\rm{inf}}}{\sigma_{p}}\right)^2 
    + \log\left(\frac{2\pi}{\sigma_p}\right)\right] \nonumber \\
 & & -\left[ \left( \frac{\Delta t-\Delta t_{{\rm pred},\infty}}{\sigma_t}\right)^2
    + \log\left(\frac{2\pi}{\sigma_t}\right)\right] \nonumber \\
 & & -\left[ \left( \frac{E_b-E_{b,{\rm pred},\infty}}{\sigma_E}\right)^2
    + \log\left(\frac{2\pi}{\sigma_E}\right)\right].
    \label{eq:lhood}
\end{eqnarray}
We then run the MC chains to refine the initial guesses, and provided convergence criteria can be achieved, we may find it possible to constrain the system parameters.

Where model predictions including the burst lightcurves are available, and the bursts are also observed at sufficient signal-to-noise to fully resolve the lightcurve observationally, we may incorporate the lightcurve comparison into our likelihood calculation.
For the purposes of comparison, we convert the model lightcurves to what a distant observer would see, following the approach described in \S\ref{subsec:simobs}.
For a given pair of observed and model lightcurves, the only parameters that affect the comparison are the source distance $d$, the anistotropy parameters $\xi_b$, $\xi_p$ (each a function of the inclination $i$), and the gravitational redshift $1+z$ (which also determines the parameter $\xi$).
We also introduce a ``nuisance'' parameter, $t_{\rm off}$, which is required to align the observed and predicted model lightcurve so as to minimise any residual differences. This parameter and the (time dilated) model timestamps $t_{i,\infty}$ are used to overlay the model predicted lightcurve onto the observed one.

Our approach is then to explore the parameter space as before, including those parameters which influence the lightcurve $(d, i, 1+z, t_{\rm off})$ to find the  set of parameters for which the comparison likelihood including the lightcurve comparison terms is maximised:
\begin{equation}
\mathcal{L} = \ldots -\sum \left[\left( \frac{F_i-F_{i,\rm{pred}}}{\sigma_{F}}\right)^2 
    + \log\left(\frac{2\pi}{\sigma_i}\right)\right], \label{eq:lccont}
\end{equation}
where $F_i$ is the burst flux at timestep $i$ within the lightcurve, and $F_{i,{\rm pred}}$ is the corresponding prediction (rescaled and shifted based on the system parameters, and presumably also interpolated onto the observation time grid).
Varying the redshift $1+z$ will allow us to obtain the best match between the model and observed lightcurve.
As each model run has been performed with a particular value of the surface gravity $g$, a particular value of $1+z$ implies in turn specific values of $M_{\rm NS}$ and $R_{\rm NS}$. 

To illustrate the capabilities of the {\sc concord} code, we carry out a single-epoch comparison of the 2007 March observation of GS~1826$-$238, as provided by \cite{gal17a}. We adopt the averaged burst lightcurve for the recurrence time of 3.53~hr, and the other parameters as listed. We use the {\tt ObservedBurst} class, which can automatically read in the information % \todo{provided it's also downloaded and put somewhere... where?}
provided it is downloaded locally and placed in the {\tt data} subdirectory of the repository.
We then incorporate a {\tt KeplerBurst} class object for the comparison. This class is designed for {\sc kepler} model results, and attributes include the accretion rate, assumed composition and surface gravity, and the average recurrence time. The lightcurve is averaged over multiple bursts in the burst train, following \cite{johnston20}. The particular model run we choose is intended to match the observed burst at low metallicity $\zcno=0.005$, which requires an accretion rate of $\dot{m}=0.1164\dot{m}_{\rm Edd}$. 

\begin{figure}[ht!]
\epsscale{1.2} % added for preprint version
\plotone{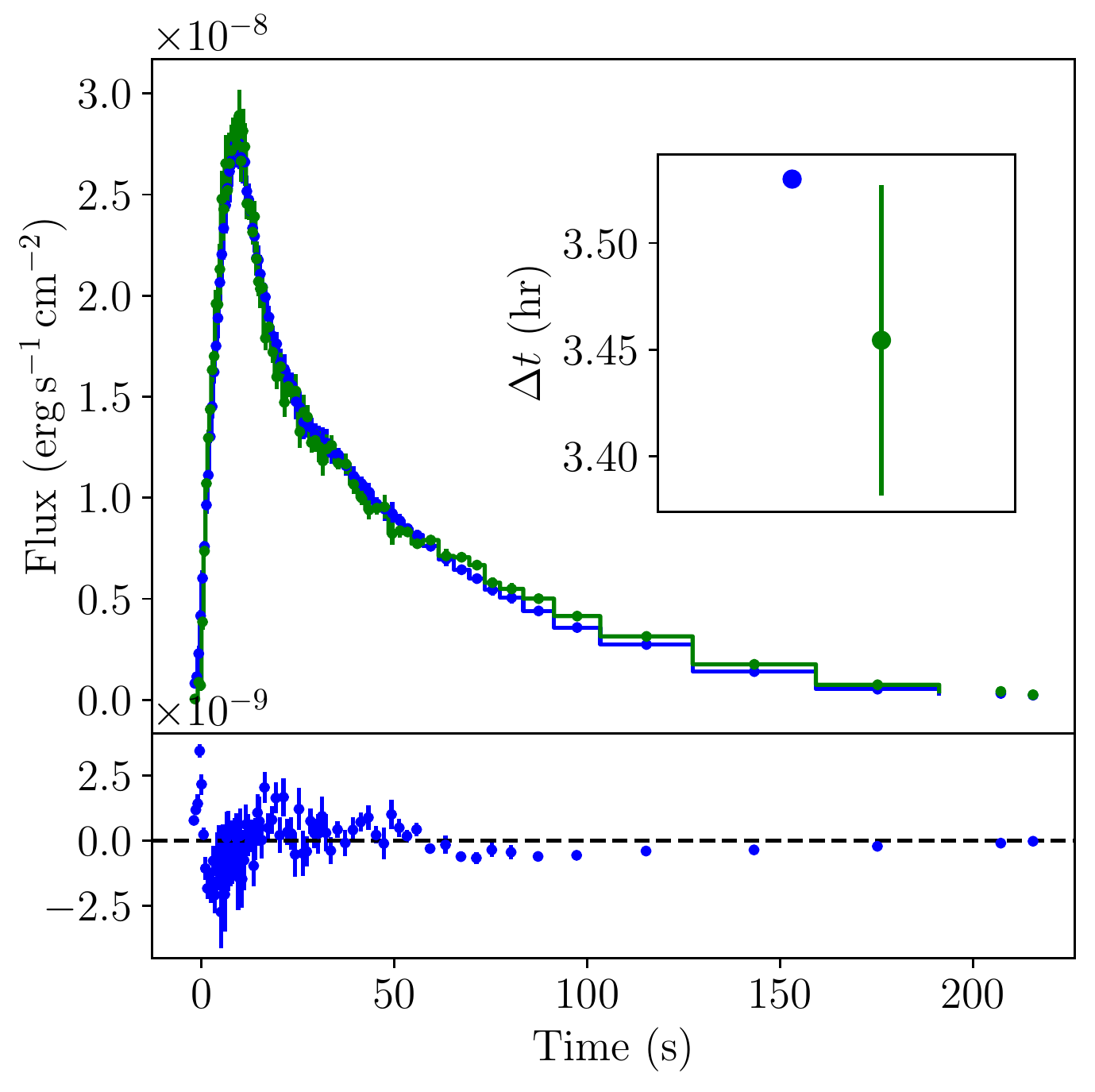}
\caption{Example model-observation lightcurve comparison with {\sc concord}. The average lightcurve of bursts observed from GS~1826$-$238 on 2007 March with \xte/PCA ({\it blue lines}) are plotted against the rescaled and interpolated model curve ({\it green lines}). The model lightcurve is transformed from the model (Newtonian) frame based on a trial set of system parameters (distance $d$, inclination $i$, and redshift $1+z$), and then interpolated onto the observed lightcurve time bins. The residuals are shown in the bottom panel; the inset also shows the comparison of the observed and predicted (redshifted) recurrence time. The average observed recurrence time ({\it blue symbol}) is 3.53~hr; in this case, the model bursts have significantly more variations in their recurrence times than the observations, as indicated by the much larger error bar ({\it green symbol}).
\label{fig:lc_comp}}
\end{figure}

We then use the {\tt compare} method of the {\tt ObservedBurst} class, with (for illustration) $d=7.5$~kpc, $i=60^\circ$, $1+z=1.31$ and a time offset of $-6.5$~s. The comparison of the re-scaled and interpolated model lightcurve is illustrated in Fig. \ref{fig:lc_comp}, including the comparison of the observed and model recurrence time. We find that the comparison is qualitatively good, and the predicted error bar uncertainty range intercepts the observed value. However, the {\tt compare} method also returns a likelihood value calculated according to equations \ref{eq:lhood} and \ref{eq:lccont}; in this case, the likelihood value is 6851.6. 
The likelihood calculation provides the capability to include this method into an MCMC calculation to estimate posterior distributions for the model parameters, $d$, $i$, $1+z$ (and the ``nuisance'' parameter $t_{\rm off}$). However, these posteriors are only relevant to the comparison with this particular model realisation, with its explicit input value for $Q_b$, $X_0$, $\zcno$ and the other model parameters.

\section{Discussion}
\label{sec:disc}

We present a new software suite, {\sc concord}, which implements functions and classes intended to facilitate analysis of thermonuclear (type-I) X-ray bursts, and hence constrain the system parameters. 
The code is written in Python and is publicly available via GitHub.

The code introduces procedures to account for commonly-encountered astrophysical uncertainties that may affect such measurements, most notably the expected anisotropy of the X-ray emission (both persistent and burst) arising from the target sources. 
The anisotropy treatment makes use of pre-calculated tables for several different disk model geometries, as simulated by \cite{he16}. The user can modify this treatment, via the addition of new ``models'' in the {\tt diskmodel} module. 

The code uses a Monte-Carlo approach for all measured quantities to propagate errors, which also offers flexibility on providing prior information on parameters that might contribute to systematic uncertainties, for example the NS mass and radius. 

We have demonstrated the utility of the code with several examples in this paper, which are also provided via a companion {\sc jupyter} notebook demonstrating how the code may be invoked.
We have also compared the results obtained with the {\sc concord} code against independent measurements from numerical simulations, to quantify their precision. 

We point out here that the calculations are only as good as the underlying assumptions, in particular the modelling of the anisotropy. For example, in \S\ref{subsec:fuelcomp} we derive the fuel composition for three epochs of burst measurements in GS~1826$-$238, the ``Clocked burster'', often used as an exemplar that behaves consistently to numerical model predictions to an unusually high level. Somewhat surprisingly, we find that the three burst epochs result in  values in $X_0$--$i$ parameter space that cannot be mutually reconciled.

We suggest that the discrepancy may arise from the assumptions about the disk structure. The default model (model ``A'') in the code assumes a thin disk that extends to the NS surface \cite[]{he16}. Such a disk is expected to be present for systems accreting at high inclination rates, and exhibiting characteristically soft  X-ray spectra for the persistent emission \cite[e.g.][]{done07}. However, during the observations of these bursts, GS~1826$-$238 exhibited instead an unusually hard X-ray spectrum, suggestive instead of truncation of the disk above the NS surface. We hypothesize that modelling including a varying disk truncation radius with accretion rate might be sufficient to reconcile the three epochs, and perhaps also resolve the discrepancy with the more detailed modelling of \cite{johnston20}.

In any case, the code offers a framework which can be developed and further adapted with (for example) additional options for anisotropy modelling, as well as other aspects which affect the observed burst properties.
We hope that this code, in combination with the large samples of burst measurements now available as well as new observations obtained in future, will permit new insights into the properties of bursting sources and hence the details of the thermonuclear burst physics.

\acknowledgments
We thank the anonymous referee, whose feedback substantially improved this
paper.
The Multi-INstrument Burst ARchive (MINBAR) project has benefited from support by the Australian Academy of Science's Scientific Visits to Europe program, and the Australian Research Council's Discovery Projects (project DP0880369) and Future Fellowship (project FT0991598) schemes. The MINBAR project has also received funding from the European Union's Horizon 2020 Programme under the AHEAD project (grant agreement no. 654215).
Parts of this research were conducted by the Australian Research Council Centre of Excellence for Gravitational Wave Discovery (OzGrav), through project number CE170100004.
This research made use of Astropy,\footnote{http://www.astropy.org} a community-developed core Python package for Astronomy \citep{astropy13, astropy18}.

\vspace{5mm}
\facilities{RXTE(PCA), INTEGRAL(JEM-X), Swift(XRT)}

\software{Astropy \citep{astropy13, astropy18}
          }

\appendix
\section{{\sc Kepler} model runs}

To validate the calculations and code in this paper (see \S\ref{subsec:models} and \S\ref{subsec:validation}) we adopt a set of {\sc Kepler} simulations originally carried out by \cite{goodwin19a}, to verify the relationship between the nuclear energy generation rate $Q_{\rm nuc}$ and the average H-fraction in the fuel layer, $\overline{X}$.
The simulation parameters are listed in Table \ref{tab:kepler}, and these data are also provided as a machine-readable table accompanying this paper. The columns are as follows: (1) run number; (2--4) input parameters to Kepler, including the accretion rate $\dot{m}$, expressed as a fraction of the Eddington rate;
the H-fraction in the accreted fuel, $X$; and the CNO mass fraction $Z$; (5) the number of bursts $n_{\rm burst}$ simulated; (6) the number of bursts averaged $n_{\rm avg}$, for the derived quantities (typically the first few bursts were discarded);  (7) the accreted mass $\Delta M$; (8) the average recurrence time $\Delta t$ and $1\sigma$ uncertainty; (9) the average burst energy $E_b$ and $1\sigma$ uncertainty; (10) the average value of the H-fraction over the fuel layer $\overline{X}$, and the $1\sigma$ uncertainty; and the average H-fraction at  the ignition point $X_{\rm ign}$.

Note that parameters (8) \& (9) are measured in the (Newtonian) model frame, and so must be adjusted for the effects of general relativity to give the equivalent values as would be measured by an observer. The surface gravity is set to $1.858\times10^{14}\ {\rm cm\,s^{-2}}$ corresponding to a $1.4\ M_\odot$ NS with a radius of 10~km (equivalent to a star of the same mass with $R_{\rm NS}=11.2$~km, accounting for general  relativity). The base flux is set consistently throughout as 0.1~MeV/nucleon; \edit1{in {\sc Kepler} this parameter is multiplied by the accretion rate to give the incoming luminosity from below the simulation zone.}

We note also that the {\sc Kepler} runs were performed without any ``preheating'' that can reduce the burn-in time \cite[see e.g.][]{johnston20}. This omission may mean that recurrence times are  overestimated by approximately 10--20\%, as otherwise the fuel layer asymptotically approaches asymptotic conditions where the temperature is maximal (and hence recurrence time is shortest).

\begin{deluxetable}{lcccccccccc}
\tablecaption{Simulated {\sc Kepler} burst train parameters \label{tab:kepler}}
\tablecolumns{11}
\tabletypesize{\scriptsize}
\tablewidth{0pt}
\tablehead{
 & & & & & & \colhead{$\Delta M$} & \colhead{$\Delta t$} & \colhead{$E_b$}  \\
\colhead{Run} & \colhead{$\dot{m}$\tablenotemark{a}} & \colhead{$X$} & \colhead{$Z$} & \colhead{$n_{\rm burst}$} &
  \colhead{$n_{\rm avg}$} & \colhead{($10^{20}$~g)} & \colhead{(hr)} & \colhead{($10^{38}\ \rm{erg}$)} &
  \colhead{$\overline{X}$} & \colhead{$X_{\rm ign}$} \\
\colhead{(1)} & \colhead{(2)} &\colhead{(3)} &\colhead{(4)} &\colhead{(5)} &\colhead{(6)} &\colhead{(7)} &\colhead{(8)} &\colhead{(9)} &\colhead{(10)} &\colhead{(11)} 
}
\startdata
1 & 0.3 & 0.2 & 0.02 & 51 & 49 & 11.924 & $1.02\pm0.03$ & $25.1\pm0.8$ & $0.160\pm0.006$ & 0.09289 \\
2 & 0.3 & 0.3 & 0.02 & 37 & 35 & 12.322 & $1.06\pm0.03$ & $33.4\pm1.3$ & $0.258\pm0.008$ & 0.16496 \\
3 & 0.3 & 0.4 & 0.02 & 36 & 34 & 11.918 & $1.03\pm0.04$ & $39\pm2$ & $0.358\pm0.008$ & 0.26395 \\
4 & 0.3 & 0.5 & 0.02 & 36 & 34 & 11.475 & $0.99\pm0.03$ & $44\pm3$ & $0.461\pm0.007$ & 0.40095 \\
5 & 0.3 & 0.6 & 0.02 & 50 & 48 & 11.352 & $0.97\pm0.02$ & $49\pm3$ & $0.563\pm0.006$ & 0.52224 \\
6 & 0.3 & 0.7 & 0.02 & 35 & 33 & 11.832 & $1.03\pm0.02$ & $58\pm5$ & $0.662\pm0.007$ & 0.62179 \\
7 & 0.3 & 0.8 & 0.02 & 24 & 22 & 12.746 & $1.12\pm0.04$ & $67\pm8$ & $0.759\pm0.009$ & 0.70960 \\
8 & 0.3 & 0.2 & 0.005 & 38 & 36 & 15.983 & $1.38\pm0.08$ & $37\pm2$ & $0.184\pm0.003$ & 0.11118 \\
9 & 0.3 & 0.3 & 0.005 & 32 & 31 & 16.427 & $1.43\pm0.05$ & $47\pm2$ & $0.283\pm0.004$ & 0.17244 \\
10 & 0.3 & 0.4 & 0.005 & 33 & 31 & 15.009 & $1.30\pm0.07$ & $51\pm5$ & $0.383\pm0.004$ & 0.29499 \\
11 & 0.3 & 0.5 & 0.005 & 36 & 35 & 13.897 & $1.20\pm0.06$ & $55\pm6$ & $0.485\pm0.004$ & 0.43364 \\
12 & 0.3 & 0.6 & 0.005 & 45 & 44 & 13.387 & $1.15\pm0.04$ & $60\pm7$ & $0.588\pm0.002$ & 0.56270 \\
13 & 0.3 & 0.7 & 0.005 & 45 & 43 & 14.239 & $1.23\pm0.03$ & $71\pm9$ & $0.6881\pm0.0019$ & 0.67223 \\
14 & 0.3 & 0.2 & 0.005 & 32 & 30 & 15.967 & $1.39\pm0.08$ & $37\pm2$ & $0.184\pm0.003$ & 0.11115 \\
15 & 0.3 & 0.2 & 0.1 & 27 & 26 & 13.082 & $1.14\pm0.06$ & $18\pm2$ & $0.05\pm0.03$ & 0.00769 \\
16 & 0.3 & 0.3 & 0.1 & 30 & 28 & 8.192 & $0.9\pm0.8$ & $18\pm4$ & $0.16\pm0.03$ & 0.02453 \\
17 & 0.3 & 0.4 & 0.1 & 68 & 67 & 8.127 & $0.692\pm0.005$ & $22\pm3$ & $0.264\pm0.017$ & 0.13451 \\
18 & 0.3 & 0.5 & 0.1 & 58 & 56 & 8.217 & $0.701\pm0.010$ & $27\pm3$ & $0.365\pm0.018$ & 0.23658 \\
19 & 0.3 & 0.6 & 0.1 & 51 & 50 & 8.067 & $0.691\pm0.008$ & $31\pm3$ & $0.467\pm0.019$ & 0.33747 \\
20 & 0.3 & 0.7 & 0.1 & 39 & 37 & 7.941 & $0.69\pm0.02$ & $35\pm3$ & $0.57\pm0.02$ & 0.43025 \\
21 & 0.3 & 0.2 & 0.1 & 26 & 25 & 13.068 & $1.14\pm0.07$ & $18\pm3$ & $0.05\pm0.03$ & 0.00800 \\
22 & 0.2 & 0.2 & 0.02 & 36 & 34 & 12.002 & $1.56\pm0.04$ & $24.0\pm1.0$ & $0.141\pm0.010$ & 0.06585 \\
23 & 0.2 & 0.3 & 0.02 & 35 & 33 & 12.469 & $1.62\pm0.04$ & $32.3\pm1.3$ & $0.239\pm0.011$ & 0.15682 \\
24 & 0.2 & 0.4 & 0.02 & 34 & 32 & 12.714 & $1.65\pm0.04$ & $39.8\pm1.7$ & $0.336\pm0.012$ & 0.22918 \\
25 & 0.2 & 0.5 & 0.02 & 32 & 31 & 12.071 & $1.57\pm0.05$ & $44\pm3$ & $0.439\pm0.011$ & 0.35049 \\
26 & 0.2 & 0.6 & 0.02 & 37 & 35 & 11.501 & $1.49\pm0.04$ & $49\pm3$ & $0.544\pm0.010$ & 0.48209 \\
27 & 0.2 & 0.7 & 0.02 & 37 & 35 & 11.668 & $1.52\pm0.04$ & $55\pm4$ & $0.645\pm0.010$ & 0.58878 \\
28 & 0.3 & 0.2 & 0.1 & 23 & 22 & 12.843 & $1.12\pm0.06$ & $18\pm4$ & $0.05\pm0.03$ & 0.00909 \\
29 & 0.2 & 0.2 & 0.005 & 24 & 22 & 17.560 & $2.30\pm0.11$ & $39.7\pm1.9$ & $0.176\pm0.005$ & 0.10829 \\
31 & 0.2 & 0.4 & 0.005 & 25 & 23 & 17.052 & $2.24\pm0.12$ & $58\pm4$ & $0.375\pm0.006$ & 0.28462 \\
32 & 0.2 & 0.5 & 0.005 & 18 & 16 & 15.771 & $2.12\pm0.07$ & $64\pm6$ & $0.475\pm0.007$ & 0.38890 \\
33 & 0.2 & 0.6 & 0.005 & 29 & 27 & 14.310 & $1.88\pm0.09$ & $64\pm8$ & $0.580\pm0.004$ & 0.53863 \\
34 & 0.2 & 0.7 & 0.005 & 29 & 27 & 14.564 & $1.91\pm0.05$ & $70\pm10$ & $0.682\pm0.004$ & 0.66101 \\
35 & 0.2 & 0.2 & 0.02 & 36 & 34 & 12.002 & $1.56\pm0.04$ & $24.0\pm1.0$ & $0.141\pm0.010$ & 0.06585 \\
36 & 0.2 & 0.2 & 0.1 & 21 & 20 & 19.768 & $2.62\pm0.06$ & $24.8\pm1.0$ & $0.03\pm0.04$ & 0.01000 \\
37 & 0.2 & 0.3 & 0.1 & 14 & 12 & 14.498 & $1.98\pm0.03$ & $23\pm4$ & $0.08\pm0.07$ & 0.02500 \\
38 & 0.2 & 0.4 & 0.1 & 40 & 2 & 4.269 & $1.03\pm0.04$ & $20\pm10$ & $0.29\pm0.11$ & 0.20000 \\
39 & 0.2 & 0.5 & 0.1 & 39 & 37 & 7.831 & $1.013\pm0.005$ & $23\pm4$ & $0.31\pm0.03$ & 0.12211 \\
40 & 0.2 & 0.6 & 0.1 & 32 & 30 & 7.930 & $1.032\pm0.013$ & $28\pm5$ & $0.41\pm0.04$ & 0.22348 \\
41 & 0.2 & 0.7 & 0.1 & 23 & 21 & 7.700 & $1.02\pm0.02$ & $32\pm4$ & $0.51\pm0.04$ & 0.32535 \\
42 & 0.3 & 0.2 & 0.1 & 60 & 4 & 10.354 & $1.12\pm0.07$ & $20\pm6$ & $0.08\pm0.07$ & 0.05000 \\
43 & 0.1 & 0.2 & 0.02 & 40 & 3 & 16.381 & $6.16\pm0.03$ & $32\pm2$ & $0.10\pm0.07$ & 0.06667 \\
44 & 0.1 & 0.3 & 0.02 & 90 & 7 & 10.722 & $3.11\pm0.09$ & $28\pm3$ & $0.19\pm0.04$ & 0.07710 \\
45 & 0.1 & 0.4 & 0.02 & 60 & 4 & 9.568 & $3.20\pm0.06$ & $37\pm4$ & $0.31\pm0.05$ & 0.20516 \\
46 & 0.1 & 0.5 & 0.02 & 40 & 3 & 8.792 & $3.28\pm0.07$ & $45\pm5$ & $0.41\pm0.06$ & 0.31122 \\
47 & 0.1 & 0.6 & 0.02 & 40 & 3 & 8.179 & $3.07\pm0.08$ & $50\pm7$ & $0.52\pm0.06$ & 0.43911 \\
48 & 0.1 & 0.7 & 0.02 & 30 & 2 & 5.767 & $2.95\pm0.05$ & $60\pm10$ & $0.65\pm0.05$ & 0.59032 \\
49 & 0.2 & 0.2 & 0.02 & 40 & 2 & 6.342 & $1.577\pm0.017$ & $26\pm2$ & $0.17\pm0.03$ & 0.13171 \\
50 & 0.1 & 0.2 & 0.005 & 11 & 9 & 18.052 & $5.11\pm0.04$ & $41.2\pm0.9$ & $0.156\pm0.016$ & 0.07394 \\
51 & 0.1 & 0.3 & 0.005 & 11 & 9 & 18.349 & $5.17\pm0.06$ & $53.4\pm0.9$ & $0.254\pm0.016$ & 0.15758 \\
52 & 0.1 & 0.4 & 0.005 & 11 & 9 & 18.583 & $5.24\pm0.07$ & $64.7\pm1.9$ & $0.353\pm0.017$ & 0.25111 \\
53 & 0.1 & 0.5 & 0.005 & 11 & 9 & 17.915 & $5.04\pm0.18$ & $72\pm4$ & $0.455\pm0.016$ & 0.36416 \\
54 & 0.1 & 0.6 & 0.005 & 12 & 10 & 15.980 & $4.4\pm0.2$ & $73\pm8$ & $0.556\pm0.015$ & 0.45273 \\
55 & 0.1 & 0.7 & 0.005 & 13 & 11 & 15.525 & $4.29\pm0.19$ & $80\pm10$ & $0.660\pm0.013$ & 0.58806 \\
56 & 0.2 & 0.2 & 0.005 & 24 & 22 & 17.560 & $2.30\pm0.11$ & $39.7\pm1.9$ & $0.176\pm0.005$ & 0.10829 \\
57 & 0.1 & 0.2 & 0.1 & 40 & 2 & 32.831 & $16.7\pm1.0$ & $70\pm20$ & $0.10\pm0.10$ & 0.10000 \\
58 & 0.1 & 0.3 & 0.1 & 60 & 4 & 28.842 & $9.71\pm0.19$ & $44\pm4$ & $<0.21$ & 0.07500 \\
59 & 0.1 & 0.4 & 0.1 & 90 & 7 & 22.018 & $6.46\pm0.05$ & $34\pm3$ & $<0.21$ & 0.05714 \\
60 & 0.1 & 0.5 & 0.1 & 11 & 9 & 18.966 & $5.38\pm0.06$ & $32\pm6$ & $0.11\pm0.14$ & 0.05555 \\
63 & 0.2 & 0.2 & 0.1 & 22 & 20 & 19.768 & $2.62\pm0.07$ & $24.7\pm1.0$ & $0.03\pm0.04$ & 0.01000 \\
\enddata
\tablenotetext{a}{Accretion rate in units of Eddington, i.e. $8.8\times10^4/(1+X)\ {\rm g\,cm^{-2}\,s^{-1}}$}
\end{deluxetable}

\bibliography{all}
\bibliographystyle{apj}

\end{document}